\documentclass[runningheads]{llncs}
\usepackage[table,dvipsnames]{xcolor}
\usepackage[misc,geometry]{ifsym}
\usepackage{amsmath}
\usepackage{newtxmath}
\usepackage{newtxtext}
\usepackage{xspace}
\usepackage{tikz}
\usepackage{xcolor}
\usepackage{paralist}
\usepackage{hyperref}
\usepackage[pass]{geometry}
\usepackage[capitalise]{cleveref}
\usepackage{enumitem}
\usepackage{multirow}
\usepackage{wrapfig}
\usepackage{listings} 
\usepackage{caption}
\usepackage{subcaption}
\usepackage{courier}

\usepackage{todonotes}

\usepackage{booktabs}

\usepackage{tikz}
\usetikzlibrary{matrix,arrows.meta,positioning,calc,decorations.pathreplacing}

\newcommand{\chck}[1]{#1}

\newcommand{\vh}[1]{\textcolor{green!50!black}{\ifmmode \text{[VH: #1]}\else [VH: #1] \fi}}
\newcommand{\ol}[1]{\textcolor{blue}{\ifmmode \text{[OL: #1]}\else [OL: #1] \fi}}
\newcommand{\js}[1]{\textcolor{red}{\ifmmode \text{[JS: #1]}\else [JS: #1] \fi}}

\newcommand{\toolname}[1]{\textsc{#1}\xspace}
\newcommand{\automatajar}{\toolname{AutomataLib}}
\newcommand{\awali}{\toolname{Awali}}
\newcommand{\lash}{\toolname{Lash}}
\newcommand{\mona}{\toolname{Mona}}
\newcommand{\mata}{\toolname{Mata}}
\newcommand{\matasim}{\toolname{Mata-Sim}}
\newcommand{\enfa}{\toolname{eNfa}}
\newcommand{\noodler}{\toolname{Z3-Noodler}}
\newcommand{\cpp}{C++\xspace}
\newcommand{\csharp}{
  {\settoheight{\dimen0}{C}C\kern-.05em \resizebox{!}{\dimen0}{\raisebox{\depth}{\#}}}\xspace}
\newcommand{\vata}{\toolname{Vata}}
\newcommand{\brics}{\toolname{Brics}}
\newcommand{\automatanet}{\toolname{Automata.net}}
\newcommand{\automatapy}{\toolname{Automata.py}}
\newcommand{\fado}{\toolname{FAdo}}

\newcommand{\code}[1]{\ensuremath{\mathtt{#1}}}
\newcommand{\ordvector}{\code{OrdVector}\xspace}
\newcommand{\symbolpost}{\code{SymbolPost}\xspace}
\newcommand{\statepost}{\code{StatePost}\xspace}
\newcommand{\pushb}{\code{push\_back}\xspace}
\newcommand{\popb}{\code{pop\_back}\xspace}
\newcommand{\ins}{\code{insert}\xspace}
\newcommand{\erase}{\code{erase}\xspace}
\newcommand{\deltastruct}{\code{Delta}\xspace}
\newcommand{\postvec}{\code{post}\xspace}

\newcommand{\post}{\mathit{post}}

\newcommand{\pow}[1]{\mathcal{P}(#1)}
\newcommand{\benchname}[1]{\textbf{\textsf{#1}}\xspace}
\newcommand{\bcsmtlib}{\benchname{b-smt}}
\newcommand{\bregexlib}{\benchname{email-filter}}
\newcommand{\bparam}{\benchname{param-inter}}
\newcommand{\bparamunion}{\benchname{param-union}}
\newcommand{\bincl}{\benchname{armc-incl}}
\newcommand{\bpres}{\benchname{lia}}
\newcommand{\bpressym}{\benchname{lia-symbolic}}
\newcommand{\bpresexpl}{\benchname{lia-explicit}}
\newcommand{\bnoodlercompl}{\benchname{noodler-compl}}
\newcommand{\bnoodlerconc}{\benchname{noodler-conc}}
\newcommand{\bnoodlerinter}{\benchname{noodler-inter}}

\newcommand{\smtbenchfont}[1]{\textit{#1}}

\usepackage{trimclip}

\makeatletter
\DeclareRobustCommand{\shortto}{%
  \mathrel{\mathpalette\short@to\relax}%
}

\DeclareRobustCommand{\shortminus}{%
  \mathrel{\mathpalette\short@minus\relax}%
}

\newcommand{\short@to}[2]{%
  \mkern2mu
  \clipbox{{.5\width} 0 0 0}{$\m@th#1\vphantom{+}{\rightarrow}$}%
}

\newcommand{\short@minus}[2]{%
  \mkern2mu
  \clipbox{{.5\width} 0 0 0}{$\m@th#1\vphantom{+}{-}$}%
}
\makeatother

\newcommand{\labeledto}[1]{{{\shortminus}\hspace{-2pt}\raisebox{0.20ex}{$\scriptstyle\hspace{1.8pt} #1\hspace{-0.28pt}\hspace{1.3pt}$}\hspace{-2.2pt}{\shortto}}}

\newcommand{\scriptlabeledto}[1]{{{\shortminus}\hspace{-1.0pt}\raisebox{0.12ex}{$\scriptscriptstyle\{ #1\hspace{-0.28pt}\}$}\hspace{-1.6pt}{\shortto}}}

\newcommand\move[3]{
\mathchoice
{#1\,\labeledto{#2}\,#3}
{#1\labeledto{#2}#3}
{#1\scriptlabeledto{#2}#3}
{#1\scriptlabeledto{#2}#3}
}

\newcommand{\concat}{\cdot}
\newcommand{\concats}{\cdots}

\newcommand{\langof}[1]{\lang(#1)}
\newcommand{\lang}{L}

\newcommand{\aut}{\mathcal{A}}
\newcommand{\but}{\mathcal{B}}

\renewcommand{\sc}[1]{#1^{\mathsf{\subseteq}}}

\newcommand{\nearzero}{$\sim$0\xspace}
\newcommand{\best}[1]{\textbf{#1}}

\newboolean{showcomments}
\setboolean{showcomments}{true}
\ifthenelse{\boolean{showcomments}}
{ \newcommand{\mynote}[3]{
    \fbox{\bfseries\sffamily\scriptsize#1}
    {\small$\ll$\textsf{\emph{\color{#3}{#2}}}$\gg$}}}
{ \newcommand{\mynote}[3]{}}
\newcommand{\shrink}[1]{}
\definecolor{pink}{rgb}{1,0.2,0.7}
\definecolor{purple}{rgb}{0.7,0,0.9}
\definecolor{darkgreen}{rgb}{0,0.5,0}

\begin{document}

\def\orcidID#1{\smash{\href{http://orcid.org/#1}{\protect\raisebox{-1.25pt}{\protect\includegraphics{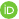}}}}}

\title{\mata: A Fast and Simple Finite Automata Library
}

\author{
David Chocholatý\orcidID{0009-0006-5614-1592} \and
Tomáš Fiedor\orcidID{0009-0009-2596-9399} \and
Vojtěch Havlena\orcidID{0000-0003-4375-7954} \and
Lukáš Holík\orcidID{0000-0001-6957-1651} \and\\
Martin Hruška\orcidID{0000-0003-2318-0940} \and
Ondřej Lengál\orcidID{0000-0002-3038-5875} \and
Juraj Síč\orcidID{0000-0001-7454-3751}
}

\authorrunning{D. Chocholatý, T. Fiedor, V. Havlena, L. Holík, M. Hruška, O. Lengál, J. Síč}

\institute{
  Faculty of Information Technology,
Brno University of Technology, Brno, Czech Republic
}

\maketitle              

\begin{abstract}
\mata is a well-engineered automata library written in \cpp that offers a unique combination of speed and simplicity. 
It is meant to serve 
in applications such as string constraint solving and reasoning about regular expressions,
and as a~reference implementation of automata algorithms.
Besides basic algorithms for (non)deterministic automata, it implements a fast simulation reduction and antichain-based language inclusion checking. The simplicity allows a straightforward access to the low-level structures, making it relatively easy to extend and modify. 
Besides the \cpp API, the library also implements a Python binding.

The library comes with a large benchmark of automata problems collected from relevant applications such as string constraint solving, regular model checking, and reasoning about regular expressions.
We show that \mata is on this benchmark significantly faster than all libraries from a wide range of automata libraries we collected.
Its usefulness in string constraint solving is demonstrated by the string solver \noodler, which is based on \mata and outperforms the state of the art in string constraint solving on many standard benchmarks.  
\end{abstract}

\section{Introduction}

We introduce a new finite automata library \mata\footnote{\url{https://github.com/VeriFIT/mata}}.
It is intended to be used in applications where automata languages are manipulated by set operations and queries, presumably in a tight loop where automata are iteratively combined together using the classical as well as special-purpose constructions. Examples are applications like
\emph{string constraint solving} algorithms such as \cite{spagheti23,ChenCHHLS23,AnthonyReplaceAll2018,Trau,Z3str3RE,norn,margus_derivatives_21}, 
processing of regular expressions \cite{cox_paper_17,dprle}, 
\emph{regular model checking} (e.g., \cite{armc04,bouajjani-antichain-08,tormc12,learningrmc17,iterating_transd03,wolperrmc98,rmc_survey04}), 
or \emph{decision procedures for logics} such as WS1S or quantified Presburger arithmetic \cite{Buchi1990,WolperB95,mona,lash}.
%
The solved problems are computationally hard, often beyond the PSPACE-completeness of basic automata problems such as language inclusion. 
Efficiency is hence a primary concern.
Achieving speed in applications requires, on one hand, fast implementation of basic automata algorithms (union, intersection, complement, minimization or size reduction, determinization, emptiness/inclusion/equivalence/membership test, parsing of regular expressions)
and, on the other hand, access to low-level primitives to implement diverse application-specific algorithms and optimizations
that often build  on a tight integration with the application environment.
Moreover, processing of regular expressions and, even more so, string constraint solving are areas of active research,  
with constantly evolving algorithms, heuristics, and optimizations. 
An automata library hence needs flexibility, extensibility,
easy access to the low-level data structures,
and ideally a low learning curve, which is important when involving students in academic research and utilizing limited resources of small research teams. 

\emph{Fast and simple} are therefore our two main requirements for the library. 
An additional third requirement is a well-engineered infrastructure and a good set of benchmarks and tests,
important for effective research and reliable deployment. 
\mata is therefore built around a data structure for the transition relation of a non-deterministic automaton that is a compromise between simplicity and speed.
It represents transitions explicitly, as triples of a sources state, a single symbol, and a target state.
This contrasts with various flavors of symbolic representation of transition relation used in advanced automata implementations in order to handle large or infinite alphabets (e.g. Unicode in processing of texts, or bit vectors in reasoning about LTL, arithmetic, or WS1S). 
However, in the applications we consider, working internally with large alphabets can essentially always be avoided by preprocessing (mainly by \emph{mintermization}, aka factorization of the alphabet). 
The simplicity of an explicit representation then seems preferable.
It allows to use a data structure specifically tailored for computing post-images of tuples and sets of states in automata algorithms:
a source state-indexed array, storing at each index the transitions from that source state in a two layered structure, with the first layer divided and ordered by symbols, and the second layer ordered by target states.
The data structure seems to be unique among the existing libraries and yields an exceptional performance.

\mata currently provides basic functionality, basic automata operations and tests, parsing of regexes and automata in a textual format, and mintermization.  
From the more advanced algorithms for working with non-deterministic automata, it implements antichain-based inclusion checking \cite{doyen-antichain-10},
and simulation-based size reduction based on the advanced algorithm of \cite{ranzato_efficient_2010,treesimulation08,lukasjirisimulation}. 
The inclusion check appears to be by a large margin the fastest implementation
available, and together with the tree automata library \vata~\cite{vata}, \mata
is the only library with an implementation of a simulation algorithm of the
second generation originating from \cite{ranzato_efficient_2010,Cece17}
(the second generation algorithms combine partition-relation pairs to
manipulate preorders that were handled explicitly by the first generation
algorithms such as~\cite{HHK95,Ilie2004}). 
%
\mata is implemented in \cpp, uses almost exclusively the STL library for its data structures, and has no external dependencies\footnote{Although, at the moment, it uses the BDD library CUDD~\cite{CUDD} in mintermisation and the regular expression parser from RE2~\cite{re2}.  The code from these projects is, however, contained within \mata. Moreover, the connection to CUDD is not tight and we plan to remove it in the future.} 
This makes it relatively easy to learn and integrate with other software projects.
It is a well-engineered project at GitHub, with modern test and quality of code assurance infrastructure.
Besides the \cpp API, it provides a Python binding for fast prototyping and easy experimenting, for instance using interactive Jupyter notebooks.

We evaluated its speed in, to our best knowledge, so far the most comprehensive comparison of automata libraries.
We compare with 7 well-known automata libraries on a large benchmark of problems from domains close to \mata's designation, mainly string constraint solving, processing regular expressions, and regular model checking. 
\mata consistently outperforms all other libraries, from several times to orders of magnitude.

That \mata is a good fit for string constraint solving is demonstrated by its central role in the string solver \noodler, which implements the algorithms of \cite{spagheti23,ChenCHHLS23}, 
and outperforms the state of the art on many standard benchmarks (see \cite{noodlertool23} for details). 

Our contributions can be summarised by the following three points:

\begin{enumerate}

\item
\mata, a fast, simple, and well-engineered automata library,  
well suited for application in string constraint solving and regex processing, in research and student projects, as well as in industrial applications.

\item
An extension of a benchmark of automata problems from string constraint solving, processing regular expressions, regular model checking, and solving arithmetic constraints.
\item
A comparison of a representative sample of well-known automata libraries against the above benchmark, demonstrating the superior performance of \mata.


\end{enumerate}

\section{Related Work}
In this overview of automata algorithms and implementations,
we focus on the technology relevant to \mata, i.e., automata used as a symbolic representation of sets of words and manipulated mainly by set operations. We omit automata technology made for other purposes, such as regular pattern matching, which concentrates on the membership test.

\paragraph{Automata techniques.}
The most textbook-like approach is to keep finite automata deterministic (the so-called DFA), which has the advantage of simple algorithms and data structures. Essentially all classical problems reduce to product construction, determinization by subset construction, final state reachability test, and minimization (by Hopcroft's~\cite{hopcroft_71}, Moore's~\cite{Moore1956}, Brzozowski's~\cite{Brzozowski1962CanonicalRE}, or Huffman's~\cite{HUFFMAN1954} algorithms). 
The obvious drawback is the susceptibility to state explosion in determinization. 

An alternative is to determinize automata only when necessary (e.g., only before complementing). 
Non-determinism may bring up to exponential savings in automata sizes and modern algorithms for \emph{nondeterministic finite automata} (NFA) can in practice avoid the exponential worst-case cost of problems like the language inclusion test. 

Namely, a major breakthrough in working with NFAs were the antichain-based algorithms for testing language universality and inclusion of NFA first introduced (to the best of our knowledge) in~\cite{TozawaH03} and later rediscovered in~\cite{dewulf_antichains_2006}.
They dramatically improve practical efficiency of the subset construction by subsumption pruning (discarding larger sets). 
They were later extended with simulation \cite{abdulla_when_2010,doyen-antichain-10} (and generalized to numerous other kinds of automata and problems).
A principally similar is the bisimulation up-to congruence technique of \cite{bonchi_checking_2013}, which optimizes the NFA language equivalence test.  Although experimental data in various works are somewhat contradictory,
the more systematic studies so far found antichain-based algorithms more efficient~\cite{chenfu_eqchecking_17,cade23}.

NFAs require more involved reduction methods than DFAs, such as those based on simulation~\cite{ranzato_efficient_2010,Cece17,HHK95,Ilie2004,lukasjirisimulation} or bisimulation~\cite{Valmari_10,piagetarjan_87,symbsim18}.
Simulation reduces significantly more but is much more costly.
The algorithms for computing simulation of the \emph{second generation}~\cite{ranzato_efficient_2010,Cece17}, which use the so-called partition-relation pairs to represent preorders on states, are practically much faster than the \emph{first generation algorithms}~\cite{HHK95,Ilie2004}. 

\newpage
\paragraph{Representations of the transition relation.}
In order to handle automata over large or infinite alphabets, such as Unicode or bit vectors, some implementations of automata represent transitions symbolically. 
Transitions may be annotated by sets of symbols represented as 
BDDs, logical formulae, intervals of numbers, etc.
The most systematic approach to this has been taken in works on  \emph{symbolic automata}~\cite{margus_rex_10,dantoni_taminimization_2016,margus_the_power17}, where the symbol predicates may be taken from any \emph{effective Boolean algebra} (essentially a~countable set closed under Boolean operations).
Some libraries, such as \toolname{Spot}~\cite{spot}, \toolname{Owl}~\cite{owl}, or \toolname{Mosel}~\cite{mosel} use BDDs to compactly represent sets of symbols on transitions.
Even more compact are the symbolic representations of the transition relation used in \mona~\cite{mona} and in the symbolic version of the tree automata library \vata~\cite{vata}, where all transitions starting at a state are represented as a single multi-terminal BDDs with the target states in the leaves (the paths represent symbols). 
Although symbolic representation may offer new optimization opportunities~\cite{dantoni_taminimization_2016} and give more generality%
, it also brings complexity and overhead. 
Adapting the known algorithms may be nontrivial \cite{dantoni_taminimization_2016,symbsim18} to the point of being a difficult unsolved problem (such as the fast computation of simulation relation of \cite{ranzato_efficient_2010,Cece17}). 
In our application area, 
working with large alphabets can mostly be avoided in preprocessing, for instance by means of a priori mintermization  
(partitioning the alphabet into groups of symbols indistinguishable from the viewpoint of the input problem). 
The simplicity and transparency of explicit representation of transitions then seems preferable. 
%

 \paragraph{Alternating automata.}
Alternating automata (AFA) received attention recently 
in the context of string solving and regex processing  \cite{fangyu_circuit_16,cox_paper_17,janku_string_2018,gange_unbounded_13}. 
They allow to keep automata operations implicit up to the point of the PSPACE-complete emptiness test, which can be solved by clever heuristics (e.g. \cite{fangyu_circuit_16,cox_paper_17,janku_string_2018,dewulf_antichains_2006,cade23,dantoni_afa_2016}).
 %
Available implementations were recently compared with selected NFA libraries \cite{cade23} and neither approach dominated. 
AFA are, however, often not a viable alternative since adapting complex algorithms from, e.g., string solving to AFA typically requires to redesign the entire algorithm from scratch (as, e.g., in \cite{janku_string_2018,fangyu_circuit_16}).  

\paragraph{String solving and SMT solvers.} 
String constraint solving is currently the primary application target of \mata. \mata is already a basis of an efficient string solver \noodler~\cite{noodlertool23} and a number of other string solvers could perhaps benefit from its performance, especially those that already use automata as a primary data structure, e.g. \cite{AnthonyComplex2019,norn,Z3str3RE,Trau}. 
Besides, SMT string constraint solvers can also be used to reason about regular properties, though the results of \cite{cade23} suggest that their efficiency is not on par with dedicated fast automata libraries.

\paragraph{Automata libraries.}
\label{sec:libraries}
We give overview of known automata libraries with a focus on those that we later include in our experimental comparison in \cref{sec:experiments}. 

The \brics~\cite{brics} automata library is often considered a baseline in comparisons. 
It implements both NFA and DFA, where each state keeps the set (implemented as a hash map) of transitions, which are represented symbolically using character ranges.
It is written in Java and relatively optimized.

The \automatanet library \cite{automatanet}, written in \csharp, implements symbolic NFA parameterized by an effective Boolean algebra.
The transition relation (as well as its inverse) are implemented as a hash map from states to the dynamic array of transitions from a~given state, each transition annotated with a predicate over the algebra.
We use it in our comparison with the algebra of BDDs.
\automatanet has been developed for a~long time and has accumulated a number of novel techniques \mbox{(e.g., an optimized minimization \cite{margus_minimization}).}

\mona \cite{mona}, written in C, is a famous optimized implementation of deterministic automata
used for deciding WS1S/WS$k$S formulae. To handle 
DFA with complex transition relations over large alphabets of bit vectors, \mona uses a~compact  fully symbolic representation of the transition relation: a single MTBDD for all transitions originating in a state, with the target states in its leaves. \mona can represent only a~DFA, hence every operation implicitly determinizes its output.

\vata \cite{vata}, written in \cpp, implements non-deterministic tree automata. 
It can be used with NFA, too as they are a special case of tree automata. It is relatively optimized and features fast implementation of the antichain-based inclusion checking \cite{bouajjani-antichain-08,tree_inclusion_11} (which for NFA boils down to the inclusion check of \cite{doyen-antichain-10}) and the second generation simulation computation algorithm of \cite{lukasjirisimulation}.

\awali \cite{awali} is a library that targets weighted automata and transducers over an arbitrary semiring.
To implement the transition relation, it keeps a vector of transitions and for each state $s$ two vectors: one keeps the indices of transitions leaving $s$ and the other one the indices of transitions entering~$s$.

\automatajar~\cite{automatajar} is a Java automata library and the basis of the automata learning framework LearnLib~\cite{learnlib}.
It focuses on DFAs and implements their transition relation as a flattened 2D matrix that maps the source state and symbol to the target state.

\automatapy~\cite{automatapy} is written in Python.
It defines the transition relation in a~liberal way, as any mapping from source states to a mapping of symbols to a~target state (DFA) or to a set of target states (NFA).

\fado~\cite{fado} is a Python library written with efficiency in mind. It uses a similar structure as \automatapy, but more specific, with the transition as a Python dictionary (a hash map), and states represented as numbers used as indices into an array.

There is a number of other automata libraries that we do not include into our comparison since they seem similar to the included ones or we were not able to use them. The C alternative of \brics \cite{libfa} and the Java implementation of symbolic NFA of~\cite{lorisjava} are in our experiment covered by \automatanet and \brics. \toolname{Alaska} \cite{alaska} contains interesting implementations of antichain-based algorithms, but is no longer maintained nor available. 
\lash \cite{lash} is a long-developed tool for arithmetic reasoning based on automata, with an efficient core automata library, written in C. 
Its transition relation is an array indexed by states, where every state is associated with a symbol-target ordered list of transitions. 
\lash uses partial symbolic representation -- it encodes symbols as sequence of binary digits. 
The comparison with \mona in \cite{Klaedtke2004} on automata benchmark originating from arithmetic problems placed its performance significantly behind \mona. 
It seems to no longer be maintained, and we were not able to run it on our benchmarks.

There is also a number of implementations of automata over infinite words, for instance \toolname{Spot} \cite{spot}, \toolname{Owl} \cite{owl}, or \toolname{Goal} \cite{goal}, which are in their nature close to the finite word automata libraries (\toolname{Spot} and \toolname{Owl} are optimized and use BDDs on transition edges similarly as \automatanet), but implement different algorithms.

\mata evolved from a prototype implementation \enfa used in the comparison of AFA emptiness checkers as a baseline implementation of classical automata \cite{cade23}. 
Surprised by its performance, we decided to turn it into a serious widely usable library. 
Current \mata is much more mature and efficient than the \enfa of \cite{cade23}.

\section{Preliminaries on Finite Automata}

\paragraph{Words and alphabets.}
%
An \emph{alphabet} is a set $\Sigma$ of \emph{symbols/letters} (usually denoted $a,b,c,\ldots$) and the set of all words over $\Sigma$ is denoted as~$\Sigma^*$.
The \emph{concatenation} of words $u$ and $v$ is denoted by $u\concat v$. 
The \emph{empty word}, the neutral element of concatenation, is denoted by~$\epsilon$ ($\epsilon \notin \Sigma$).

\newcommand{\aore}{a}
\paragraph{Finite automata.}
A~\emph{(nondeterministic) finite automaton (NFA)} over an alphabet $\Sigma$ is a tuple
$\aut= (Q,\post,I,F)$
 where $Q$ is a finite set of \emph{states},
$\post\colon Q\times(\Sigma\cup\{\epsilon\}) \to 2^Q$ is a \emph{symbol-post function}, $I\subseteq Q$ is the set of \emph{initial states}, and $F\subseteq Q$
is the set of \emph{final states}.
A~\emph{run} of~$\aut$ over a~word~$w \in \Sigma^*$ is
a~sequence
 $p_0\aore_1p_1\aore_2
  \ldots 
 \aore_n p_n$  
where for all $1\leq i \leq n$ it holds that $\aore_i \in \Sigma \cup
\{\epsilon\}$, $p_i \in \post(p_{i-1}, \aore_i)$, and $w = \aore_1
\concat \aore_2 \concats \aore_n$. 
The run is \emph{accepting} if $p_0 \in I$ and $p_n\in F$, and the language
$\langof{\aut}$ of $\aut$ is the set of all words for which $\aut$ has an accepting run.
$\aut$ is called \emph{deterministic (DFA)} if $|I| \leq 1$, $|\post(q,\epsilon)| = 0$, and $|\post(q,a)| \leq 1$ for each $q\in Q$ and $a \in \Sigma$.
A state is \emph{useful} if it belongs to some accepting run, else it is \emph{useless}. 
An automaton with no useless states is \emph{trimmed}.
A state is \emph{reachable} if it appears on a run starting at an initial state. 
In \mata, we further use $\post(q) = \{ (a, \post(q, a)) \mid \post(q,a) \neq \emptyset \}$ to denote 
the \emph{state-post} of~$q$.
We call symbol-post and state-post the \emph{post-image functions}. We also use $\move{q}{a}{p}$ where $p \in \post(q,a)$ to denote \emph{transitions}.
The set of all transitions of $\aut$ is called the \emph{transition relation} of $\aut$ and we denote it by $\Delta$.
%

\paragraph{Automata operations.}
In this paragraph we assume automata without $\epsilon$ transitions.
The \emph{subset construction} generates from~$\aut$ the DFA 
$(\sc Q,\sc{\post\!}, \sc I, \sc F)$ 
where $\sc Q = \pow Q$,
$\sc I = \{I\}$, 
$\sc F = \{S \in \sc Q\mid S\cap F\neq\emptyset\}$,
and where $\sc{\post\!}(S,a) = \bigcup_{s \in S} \post(s, a)$. The automaton for \emph{complement} is obtained from it by complementing $\sc F$, i.e., the set of final 
states is given as $\sc Q \setminus \sc F$. 
The \emph{intersection} of two automata $\aut_1 = (Q_1,\post_1,I_1,F_1)$ and $\aut_2= (Q_2,\post_2,I_2,F_2)$  is implemented by their product $(Q_1\times Q_2,\post^{{\times}},I_1\times I_2,F_1\times F_2)$ where $\post^{{\times}}((q,r),a) = \post_1(q,a) \times \post_2(r,a)$.
A sensible implementation of course only computes the reachable parts of the product and the subset construction. 
The \emph{union} $L(\aut_1) \cup L(\aut_2)$ is obtained by disjointly uniting all components of~$\aut_1$ and $\aut_2$. Similarly, the \emph{concatenation} 
$L(\aut_1) . L(\aut_2)$ is the automaton $(Q_1\uplus Q_2,\post_1 \uplus\post_2 \uplus \post',I_1,F')$ where $\uplus$ denotes the disjoint union, $\post'(q,a) = \{ r \mid  q\in F_1 \land \exists s\in I_2\colon r \in \post_2(s,a) \}$ is the 
connecting symbol-post and $F'$ is $F_2$ if $I_2 \cap F_2 = \emptyset$ and $F_1 \cup F_2$ otherwise  (this construction avoids introducing $\epsilon$-transitions).
Note that we omit superscript of symbol-post function when it is clear from the context.

\newpage
\section{The Architecture of \mata}

%

We explain in this section the implementation techniques that make \mata efficient on a wide range of automata operations.

\subsection{Automata Representation}

States and transition symbols are unsigned integers (starting from 0). This makes it easy to store information about them in a state-/symbol-indexed vectors. 
A frequently used low-level data structure is \ordvector, a set of ordered elements implemented as an ordered array (with \code{std{::}vector} as the underlying data structure).
It has constant time addition and removal of the largest element (\pushb and \popb), linear union, intersection, and difference (by variants of merging), good memory locality and fast iteration through elements, logarithmic lookup (by binary search), but a slow insertion and removal (\ins and \erase) at other than the last position, as the elements on the right of the modified position must be shifted.
Many \mata algorithms utilize the constant time handling of the largest element in, e.g., synchronized traversal of multiple \ordvector containers.
Initial and final states are kept in sparse sets~\cite{sparseset93}, with fast iteration through elements and constant lookup, insertion, and removal.

\begin{figure}[t]
\begin{center}
\definecolor{color1}{RGB}{54,174,124}
\definecolor{color2}{RGB}{24,116,152}
\begin{tikzpicture}[
    circ/.style={draw,circle,inner sep=0pt,minimum size=2pt,fill},
    arr/.style={->,thick,>=stealth},
    type/.style={color2,dashed,thick},
]

\matrix[
    matrix of nodes,
    nodes={draw, minimum size=5mm},
    row sep=0.5mm,
    nodes in empty cells,
    row 1/.style={nodes={draw=none, fill=none, minimum size=5mm}},
] (delta) {
0 & 1 & 2 & 3 & 4 & 5 & 6 & 7 & 8 & 9\\
  &   &   &   &   &   &   &   &   &  \\
};
\node[below = 0cm of delta-2-8,xshift=4.5mm] {\code{std{::}vector{<}StatePost{>}}};
\draw[decoration={brace,amplitude=10pt}, decorate, color1, thick] (delta-1-1.north west) -- (delta-1-10.north east) node [above = 10pt, pos=0.5] {Source states};
\draw[type] plot [smooth,tension=2] coordinates {($(delta.west)+(-2,-5)$) ($(delta)+(0,1.5)$) ($(delta.east)+(2,-5)$)};
\node[right = 1cm of delta,color2] {\code{Delta}};

\matrix[
    matrix of nodes,
    nodes={draw, minimum size=5mm, anchor=center, 
    },
    below = 1.3cm of delta-2-6
] (statepost) {
$a$ & $c$ & $e$ & $r$ & $x$ & $\epsilon$ \\
};
\draw[arr] ($(delta-2-5.north west)!0.5!(delta-2-5.south east)$) node[circ]{} .. controls +(0,-.7) and +(0,0.7) .. (statepost-1-1.north west);
\node[below = 0cm of statepost-1-5] {\ordvector{}\code{{<}SymbolPost{>}}};
\node[above = 0cm of statepost-1-4, color1] {Transition symbols};
\draw[type] plot [smooth,tension=2] coordinates {($(statepost.west)+(-1.5,-2.5)$) ($(statepost)+(0,1)$) ($(statepost.east)+(1.5,-2.5)$)};
\node[right = 0.8cm of statepost,color2] {\code{StatePost}};

\matrix[
    matrix of nodes,
    nodes={draw, minimum size=5mm, anchor=center, 
    },
    below = 1.3cm of statepost-1-2
] (symbolpost1) {
1 & 3 & 5 & 6\\
};
\draw[arr] ($(statepost-1-2.north west)!0.5!(statepost-1-2.south east)+(0.17,0)$) node[circ]{} .. controls +(0,-.7) and +(0,0.7) .. (symbolpost1-1-1.north west);
\node[below = 0cm of symbolpost1-1-3] {\ordvector{}\code{{<}State{>}}};
\node[above = 0cm of symbolpost1-1-3, color1] {Target states};
\draw[type] plot [smooth, tension=1.1] coordinates {($(symbolpost1.south west)+(-0.2,0)$) ($(statepost-1-2)+(0,0.15)$) ($(symbolpost1.south east)+(0.2,0)$)};
\node[right = 0.2cm of symbolpost1,color2] {\code{SymbolPost}};

\end{tikzpicture}
\end{center}
\vspace{-3mm}
\caption{The transition relation.}
\label{fig:delta}
\vspace{-1mm}
\end{figure}
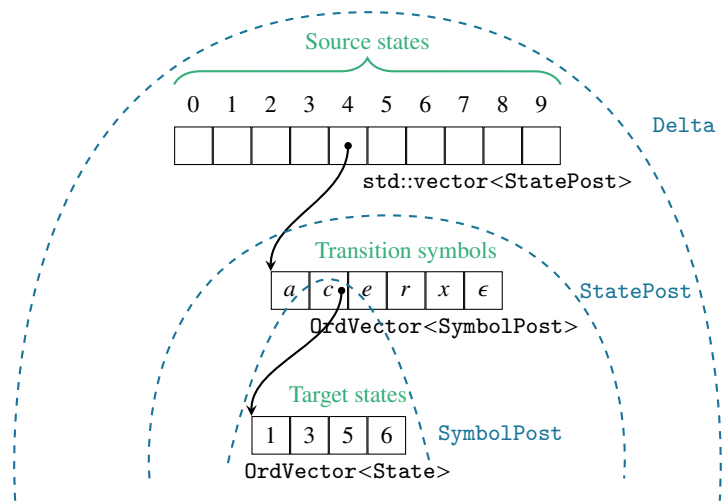

\paragraph{Data structure for the transition relation.}

The main determinant of \mata is its three-layered data structure \deltastruct
for the transition relation.
It is implemented as a vector \postvec where, for every state $q$, $\code{\postvec[}q\code{]}$ is of the type \statepost, representing $\post(q)$ as an \ordvector of objects of the type \symbolpost, each in turn representing one $\post(q,a)$ by storing the symbol $a$ and an \ordvector of the target states.
The \symbolpost{}s in \ordvector are ordered by their symbols%
\footnote{
\mata supports $\epsilon$-tran\-si\-tions and some operations can work with them internally.
We represent $\epsilon$ as the symbol with the highest possible number, hence \symbolpost with $\epsilon$ is always the last one in the vectors of \symbolpost{}s in \deltastruct. 
The~$\epsilon$ is therefore easy to be accessed in, e.g., $\epsilon$-transition elimination.
Some operations also support several $\epsilon$-like symbols (e.g.,
$\epsilon_1$, $\epsilon_2$, \ldots), which are convenient in some algorithms in
string solving~\cite{spagheti23,ChenCHHLS23} or can play a role of different
synchronization symbols, etc.
}.
A~visualization of \deltastruct is shown in \cref{fig:delta}.

The weak point of \deltastruct is inherited from \ordvector: slow \ins or
\erase of a specific transition (these operations are, however, used scarcely
in the considered scenarios).
Its strength is mainly fast iteration through the post-image of a~state, of
a~pair of states in the product construction, and of a~set of states in the
subset construction.

\subsection{Automata Operations}


\paragraph{Generating post-images in subset construction.}
In the subset construction, each iteration through $\post(S)$ for a set of states $S$ is keeping an array of iterators, one into each $\post(q)$ for all $q\in S$. Every iteration shifts the iterators to the right, to $\post(q,b)$ where $b$ is the closest from above to the current global minimal symbol~$a$, and returns $\post(S,a)$ as the union of all $\post(q,a)$'s pointed to by the iterators.
No searching in vectors is needed. The entire iteration through all $\post(S)$'s makes the iterators in the \symbolpost{}s traverse their respective vectors only once.

Constructing the transitions leading from $S$ while iterating through $\post(S)$ is done by appending to \ordvector{}s, without a need to insert at internal positions of vectors. The iteration through the \symbolpost{}s is ordered by symbol, hence each newly created transition from the macrostate $S$ has a larger symbol than all the previously created ones. The symbol-post therefore belongs at the end of the \ordvector of symbol-posts of $\post(S)$, where it is \pushb{}ed.
Since the resulting automaton is deterministic, the vectors of targets are singletons, and their creation does not require \ins either.

\paragraph{Generating post-images in product construction.}
Similarly as in the subset construction above, iterating through $\post((q,r))$ in the product construction is done by synchronous iteration through $\post(q)$ and $\post(r)$ from the smallest common symbol to the largest.
In each step, the iteration returns the Cartesian product of the targets in the symbol-posts.
%
Unlike the subset construction, adding the corresponding transitions from $(q,r)$ to the product automaton sometimes does need an \ins into the vector of targets.
It is however not that frequent:
Newly discovered product states are assigned the so far highest numbers,
so these are added to the target vectors by \pushb.
The \ins may hence be needed only when creating a non-deterministic transition to a state discovered earlier.

\paragraph{Storing sets and pairs of states in the subset and product construction.}
\ordvector is also used to map generated sets in the subset to the identities of generated states. The map uses a hash table (\code{std{::}unordered\_map}) where values are \ordvector{}s.
The product construction uses either a two-dimensional array to map pairs of states to product states (for smaller automata) or a vector \code{pro\_map} of hash tables, where the identity of the product state $(q,r)$ is found in the hash map \code{prod\_map[q]} under the key $r$.


\paragraph{Emptiness test and trimming.}
Emptiness test and trimming are used frequently and must be fast.
\mata's emptiness test is just a~state space exploration that utilizes the fast iteration through post-images of a state.

Trimming  consists of two steps: (1)~identification of useful states and (2)~removal of states that are not useful.
Identification of useful states must, besides forward exploration to identify reachable states, identify states that reach a final state. A naive solution would be a backward exploration from final states.
\deltastruct is, however, not well suited for backward search and although reverting it is doable, its cost is not negligible either.
We therefore use a smarter solution, which uses a simplification of the
non-recursive Tarjan's algorithm~\cite{Tarjan71} to discover strongly connected
components (SCCs).
Tarjan's algorithm is essentially a depth-first exploration augmented to identify the SCCs. To identify useful states, on finding an SCC with a final state, we mark the entire SCC as useful together with all states on the path to that SCC, which is readily stored on the depth-first search stack. The cost of computing useful states is then similar to the cost of a single depth-first exploration, which is indeed negligible.

Removal of useless states then needs to be done in a \deltastruct-friendly way.
The naive approach that removes useless states and transitions incident with
them one by one would be extremely slow due to the need of searching and
calling \erase in the \ordvector{}s of \deltastruct.
Instead, we perform the whole removal and related operations in a~single pass through \deltastruct.
Before the pass begins, first, we create a~map \code{renaming} mapping each
useful state to its new name (the trimming also renames the states in order to have
the remaining states form a~consecutive sequence).
During the pass, the following operations need to be performed:
\begin{inparaenum}[(i)]
  \item  in the outermost loop, each useful state~$q$ in \deltastruct is moved
    to index \code{renaming[}$q$\code{]},
  \item  in every vector of target states, each useful target is moved to the
    left in the target vector by that many positions, as there were smaller
    useless states before it, and
  \item  while doing that, the target state~$q$ is renamed to
    \code{renaming[}$q$\code{]}.
\end{inparaenum}
%


\paragraph{Union and concatenation.}
\mata is relatively slow in operations that copy or create large parts of automata, such as non-deterministic union or concatenation, or simple copying of an automaton.
This is perhaps due to the imperfect memory locality (the three layers of vectors in \deltastruct) and the need to copy every single transition (unlike, e.g., symbolic automata with BDDs on transitions, where the BDDs may be shared).
\mata has, however, in-place variants of union and concatenation, which do not copy \deltastruct, but only append the \postvec vectors and rename the target states in the appended part, which is fast. The price for the speed is the loss of the original automata, but they are in many use cases not needed (as, e.g., in inductive constructions of automata from regular expressions or formulae).

\paragraph{Antichain-based inclusion checking.}
\mata implements the antichain-based inclusion checking of~\cite{doyen-antichain-10}. Given the inclusion problem $L(\aut) \subseteq L(\but)$, the algorithm explores the space of the product of $\aut$ and the subset construction on $\but$, consisting of pairs $(q,S)$ with $q$ being a~state of $\aut$ and $S$ being a~set of states of $\but$. In particular, it searches, on the fly, for a~reachable pair $(q,S)$ with a final $q$ and a non-final $S$, which would be a~witness non-inclusion.
The algorithm optimizes the search by \emph{subsumption pruning}---discarding states $(q,S)$ if another $(q,S')$ with $S\subseteq S'$ has been found.
Our implementation uses the infrastructure for computing post-images of product and subset construction discussed above. The reached pairs $(q,S)$ are stored in a state-$q$-indexed vector $\code{incl\_map}$ of collections of sets $S$. The sets are again represented as \ordvector{}s.
On reaching a pair $(q,S)$, all sets $S'$ stored in
$\code{incl\_map[}q\code{]}$ are tested for inclusion with~$S$.
If $S\supseteq S'$, then $S$ is dropped, and if $S\subseteq S'$, then $S'$ is
removed from $\code{incl\_map[}q\code{]}$ (as well as other sets $S''$ such
that $S \subseteq S''$) and $S$ is added to $\code{incl\_map[}q\code{]}$.
A large speed-up is sometimes obtained by prioritizing exploration of pairs $(q,S)$ with~$S$ being of a~small size. A smaller set means a better chance to subsume other pairs, to reach a witness of non-inclusion, and to generate other pairs with small sets. The algorithm then explores a much smaller state space.

\paragraph{Simulation.}
\mata uses an implementation of a~fast algorithms for computing simulation,
namely, the algorithm from~\cite{ranzato_efficient_2010}, which was adapted
from Kripke structures to automata in~\cite{treesimulation08}, and later further
optimized in~\cite{lukasjirisimulation}.
The implementation originates in \vata ~\cite{vata}.


\paragraph{Low-level API.}
The API of \mata contains an interface for accessing the most low-level
features needed to implement algorithms in the style described above.
For instance, the API provides iterators over transitions of~$\Delta$ in the
form of triples $\move q a r$, iterators through \emph{moves} (pairs $(a,r)$
such that $\move q a r \in \Delta$) of a state $q$, or generic
\emph{synchronized iterators}, which allow a~simultaneous iteration in a set of
vectors used in union and in computing the post-image in the product and subset
construction.
Since the main data structures are not complicated and have simple invariants, programming with them on the low level is possible even for an outsider.
This low-level \mata API is, for instance, used in
the string solver \noodler.
\cite{noodlertool23} presents a detailed comparison
of \noodler with the state of the art in string solving.
Its exceptional performance on regex and word equation-heavy constraints is to
a large degree due to \mata. 

\section{Infrastructure of \mata}
\label{sec:infrastructure}




\mata comes with the following tools and features to make using, developing, and extending it convenient.

\lstset{basicstyle=\fontfamily{pcr}\fontsize{6}{6}\selectfont}
\begin{figure}[t]
    \centering
    \begin{subfigure}[b]{0.75\textwidth}

        \begin{lstlisting}[language=python]
from libmata import nfa, alphabets, parser, plotting
aut1 = parser.from_regex('((a+b)*a)*')
aut2 = parser.from_regex('aab*')
con_aut = nfa.nfa.concatenate(aut1, aut2).trim()
plotting.store()['alphabet'] = \
    alphabets.OnTheFlyAlphabet.from_symbol_map({'a':97, 'b':98})
e_h = [
    (lambda aut, e: e.symbol == 98, {'color':'black'}),
    (lambda aut, e: e.symbol == 97, {'style':'dashed','color':'black'})
]
n_h = [
    (lambda aut, q: q in aut.final_states,
        {'color':'red','fillcolor':'red'}),
    (lambda aut, q: q in aut.initial_states,
        {'color': 'orange', 'fillcolor': 'orange'}),
]
plotting.plot(con_aut, with_scc=True,
              node_highlight=n_h, edge_highlight=e_h)
        \end{lstlisting}
        \vspace*{-2mm}
        \caption{An example of using \mata from Python.}
        \label{fig:vis-code}
    \end{subfigure}
    \hspace{-6mm}
    \begin{subfigure}[b]{0.28\textwidth}
        \centering
        \includegraphics[width=\textwidth]{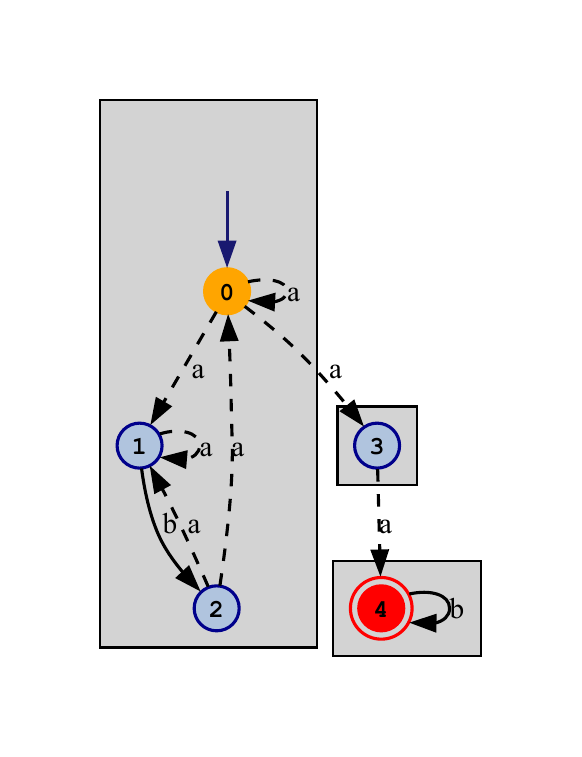}
        \vspace{-5mm}
        \caption{The output.}
        \label{fig:vis-aut}
    \end{subfigure}
       \caption{An example of a Python interface for \mata. The code (\subref{fig:vis-code}) loads
       automata from regular expressions (\texttt{a}, \texttt{b} are transition symbols; \texttt{*}, and \texttt{+} represent iterations: 0 or more, and 1 or more, respectively), concatenates them, and displays the trimmed concatenation using
       the conditional formatting with the output in (\subref{fig:vis-aut}).}
       \label{fig:python-visualisation}
\end{figure}

\paragraph{Python interface.}
\mata provides an easy-to-use Python interface, making it a full-fledged automata library for Python projects.
It is available on the official Python package
repository\footnote{\url{https://pypi.org/project/libmata/}} and can be
installed easily using the \texttt{pip} package manager:
%
\begin{center}
  \tt \$ pip install libmata
\end{center}
%
An example of using the \mata Python binding is shown in \cref{fig:python-visualisation}.
The interface is implemented using the optimizing static compiler Cython wrapping the \cpp \mata calls and
covers all important parts of the \cpp functionality.
This low-level interaction with the optimized \cpp code keeps the Python code
fast.
To show the capabilities of the
interface and to provide material for easy
onboarding, \mata also contains several Jupyter notebooks with examples of
how to use it.

\paragraph{\code{.mata} format and parsing.}
\begin{wrapfigure}[15]{r}{45mm}
  \lstset{basicstyle=\fontfamily{pcr}\scriptsize}
\vspace{-2mm}
  \begin{minipage}{\linewidth}
      \centering\captionsetup[subfigure]{justification=centering}
      \begin{lstlisting}[language=python]
@NFA-explicit
%Initial q0 q1
%Final q1
q0 a48 q1
q0 a52 q1
q1 a48 q1
      \end{lstlisting}
      \vspace{-2mm}
      \subcaption{NFA with explicit alphabet.}
      \label{fig:mata-explicit}\par\vfill
      \begin{lstlisting}[language=python]
@NFA-bits
%Initial q1
%Final q2 q1 q0
q0 ((!a0 | !a1) & a2) q2
q1 (a0 & a1 & !a2) q0
q2 ((a0 & a1) | a2) q1
      \end{lstlisting}
      \vspace{-2mm}
      \subcaption{NFA with symbolic alphabet.}
      \label{fig:mata-bits}
  \end{minipage}
     \caption{Examples of NFAs in the \code{.mata} format.}
     \label{fig:mata-format}
\end{wrapfigure}
\mata brings its own automata format. The main features of the format
are extensibility to cover various types of automata, human-readability, yet
still high level of compactness.
Each \code{.mata} file consists of automata definitions.
The first line of the definition describes the type of the automaton, together
with the alphabet.
The format supports both explicit and symbolic (bit vector) alphabets.
For a~symbolic alphabet, symbols are encoded as formulae over atomic propositions, where
the parser of \code{.mata} implements \emph{mintermization} (partitioning the alphabet
into groups of symbols indistinguishable from the viewpoint of the input
problem), which transforms it into an explicit alphabet with the symbols
representing the minterms.
The following lines contain a~sequence of key-values statements that set particular
traits of the automaton, such as initial or final states. The rest of the definition
is a list of transitions. Examples of automata in \code{.mata} format are shown in
\cref{fig:mata-format}.

Other than the introduced format, \mata can also parse automata from regular
expressions using the parser from the regex matcher RE2~\cite{re2}.
This means that \mata can handle even complex syntax used in real-world regular expressions.

\paragraph{Continuous integration.}
%
We implement continuous integration via GitHub Actions.
In particular, actions automatically build the library including the Python binding on MacOS and Ubuntu, check for warnings, code
quality and run unit tests together with the code coverage. The actions are triggered after each commit, and the checks are mandatory
for merging branches to the main branch, and can also be run locally. 

\vspace{-0.0mm}
\section{Experimental Evaluation}
\label{sec:experiments}
\vspace{-0.0mm}

\begin{figure}[t]
\captionsetup{font=small}
\begin{center}
\hspace{-3mm}
\includegraphics[width=\linewidth]{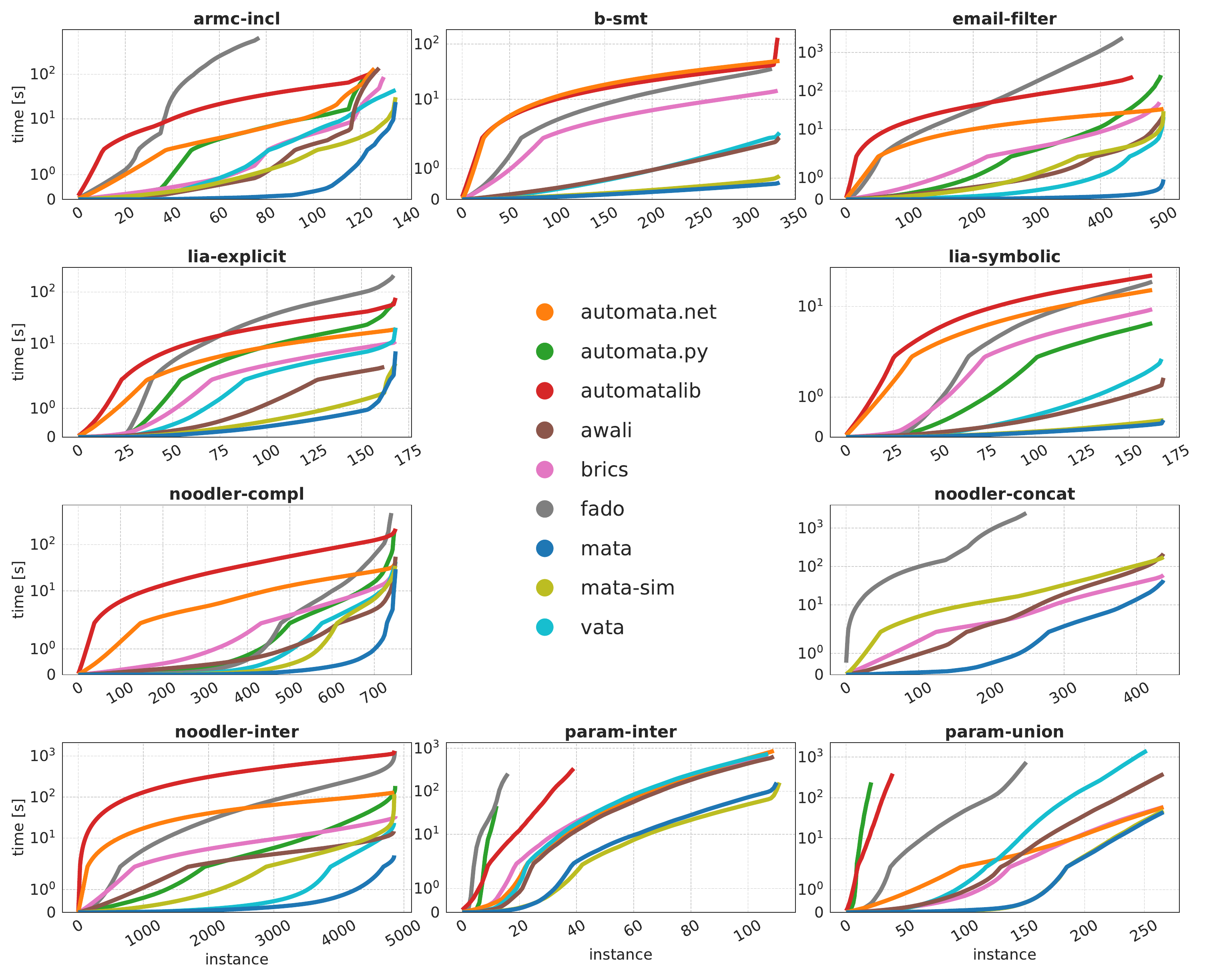}
\caption{Cactus plot showing cumulative run time per benchmark. The time axis is logarithmic.}
\label{fig:benchmarks}
\end{center}
\vspace{-4mm}
\end{figure}

We compared \mata against 7 selected libraries discussed in
\cref{sec:libraries}: 
\vata \cite{vata}, \brics \cite{brics}, \awali
\cite{awali}, \automatanet \cite{automatanet}, \automatajar \cite{automatajar},
\fado~\cite{fado}, and \automatapy~\cite{automatapy},%
\footnote{We also tried to compare with \mona, but using it as a 
standalone library that would parse automata in our format turned to be problematic. We were getting many inconsistent results and so we decided to drop it from the comparison.}
 on a benchmark of basic
automata problems from string constraint solving, reasoning about regular
expressions, regular model checking, and a~few examples from solving arithmetic
formulae. Most of the benchmark problems are taken from earlier works
\cite{cade23,dantoni_afa_2016,gange_unbounded_13,cox_paper_17}, but we
added new problems from string constraint solving and solving quantified linear
integer arithmetic (LIA).

We mainly aim to demonstrate the efficiency of the basic data structures and
implementation techniques of \mata. This is best seen on standard constructions,
where all libraries implement the same high-level algorithm, such as
product, subset construction, or reachability test within complementation,
intersection, emptiness test, etc.
We then also showcase the efficiency of more advanced algorithms implemented only in \mata and \vata,
the antichain-based inclusion test and
simulation reduction.


\begin{table}[t]
\begin{center}
\caption{Statistics for the benchmarks. We list the number of timeouts (TO), average time on solved instances (\textit{Avg}), median time over all instances (\textit{Med}), and standard deviation over solved instances (\textit{Std}), with the best values in \textbf{bold}. The times are in milliseconds unless seconds are explicitly stated. We use \nearzero to denote a~value close to zero. }
\label{table:benchmarks}
\newcolumntype{g}{>{\columncolor{gray!30}}r}
\newcolumntype{h}{>{\columncolor{gray!30}}c}
\resizebox{\textwidth}{!}{%
\begin{tabular}{l gggg rrrr gggg rrrr gggg}
 \toprule
 & \multicolumn{4}{h}{\bincl (136)}       & \multicolumn{4}{c}{\bcsmtlib (384)}   & \multicolumn{4}{h}{\bregexlib (500)}     & \multicolumn{4}{c}{\bpresexpl (169)}   & \multicolumn{4}{h}{\bpressym (169)}   \\\cmidrule(lr){2-5}\cmidrule(lr){6-9}\cmidrule(lr){10-13}\cmidrule(lr){14-17}\cmidrule(lr){18-21}
& \multicolumn{1}{h}{\textit{TO}} & \multicolumn{1}{h}{\textit{Avg}} & \multicolumn{1}{h}{\textit{Med}} & \multicolumn{1}{h}{\textit{Std}} & \multicolumn{1}{c}{\textit{TO}} & \multicolumn{1}{c}{\textit{Avg}} & \multicolumn{1}{c}{\textit{Med}} & \multicolumn{1}{c}{\textit{Std}} & \multicolumn{1}{h}{\textit{TO}} & \multicolumn{1}{h}{\textit{Avg}} & \multicolumn{1}{h}{\textit{Med}} & \multicolumn{1}{h}{\textit{Std}} & \multicolumn{1}{c}{\textit{TO}} & \multicolumn{1}{c}{\textit{Avg}} & \multicolumn{1}{c}{\textit{Med}} & \multicolumn{1}{c}{\textit{Std}} & \multicolumn{1}{h}{\textit{TO}} & \multicolumn{1}{h}{\textit{Avg}} & \multicolumn{1}{h}{\textit{Med}} & \multicolumn{1}{h}{\textit{Std}}\\
\midrule
\rowcolor{GreenYellow}
\mata             &\best 0 & \best{174} & \best 2 & 1\,s &\best 0 & \best 1 & \best 1 & 1 &\best 0 & \best 1 & \best \nearzero & 9 &\best 0 & 42 & \best 6 & 356 &\best 0 & \best 2 & \best 2 & 6\\
\awali            & 7 & 1\,s & 17 & 3\,s &\best 0 & 6 & 6 & 4 &\best 0 & 46 & 4 & 162 & 6 & \best{21} & 21 & 16 &\best 0 & 8 & 7 & 14\\
\vata             &\best 0 & 324 & 43 & 577 &\best 0 & 7 & 7 & 10 &\best 0 & 42 & 2 & 322 &\best 0 & 121 & 51 & 671 & 1 & 11 & 10 & 11\\
\automatanet      & 9 & 1\,s & 125 & 3\,s &\best 0 & 148 & 153 & 30 &\best 0 & 69 & 66 & 30 &\best 0 & 113 & 117 & 49 & 6 & 103 & 107 & 33\\
\brics            & 5 & 659 & 34 & 2\,s & 4 & 43 & 43 & 19 & 6 & 103 & 17 & 280 &\best 0 & 66 & 62 & 63 & 6 & 55 & 60 & 33\\
\automatajar      & 10 & 843 & 669 & 1\,s & 7 & 390 & 126 & 3\,s & 48 & 516 & 390 & 521 &\best 0 & 458 & 285 & 1\,s & 6 & 164 & 173 & 52\\
\fado             & 58 & 8\,s & 22\,s & 10\,s & 9 & 109 & 112 & 67 & 64 & 6\,s & 1\,s & 11\,s & 1 & 1\,s & 727 & 2\,s & 6 & 135 & 149 & 105\\
\automatapy       & 10 & 913 & 133 & 3\,s & 334 & 24 & TO & 15 & 4 & 520 & 19 & 2\,s & 1 & 372 & 167 & 894 & 6 & 35 & 35 & 25\\
\bottomrule

\end{tabular}
}

\resizebox{\textwidth}{!}{%
\begin{tabular}{l gggg rrrr gggg rrrr gggg}
 \toprule
 & \multicolumn{4}{h}{\bnoodlercompl (751)}       & \multicolumn{4}{c}{\bnoodlerconc (438)}                   & \multicolumn{4}{h}{\bnoodlerinter (4872)}   & \multicolumn{4}{c}{\bparam (267)}   & \multicolumn{4}{h}{\bparamunion (267)}   \\\cmidrule(lr){2-5}\cmidrule(lr){6-9}\cmidrule(lr){10-13}\cmidrule(lr){14-17}\cmidrule(lr){18-21}
& \multicolumn{1}{h}{\textit{TO}} & \multicolumn{1}{h}{\textit{Avg}} & \multicolumn{1}{h}{\textit{Med}} & \multicolumn{1}{h}{\textit{Std}} & \multicolumn{1}{c}{\textit{TO}} & \multicolumn{1}{c}{\textit{Avg}} & \multicolumn{1}{c}{\textit{Med}} & \multicolumn{1}{c}{\textit{Std}} & \multicolumn{1}{h}{\textit{TO}} & \multicolumn{1}{h}{\textit{Avg}} & \multicolumn{1}{h}{\textit{Med}} & \multicolumn{1}{h}{\textit{Std}} & \multicolumn{1}{c}{\textit{TO}} & \multicolumn{1}{c}{\textit{Avg}} & \multicolumn{1}{c}{\textit{Med}} & \multicolumn{1}{c}{\textit{Std}} & \multicolumn{1}{h}{\textit{TO}} & \multicolumn{1}{h}{\textit{Avg}} & \multicolumn{1}{h}{\textit{Med}} & \multicolumn{1}{h}{\textit{Std}}\\
\midrule
\rowcolor{GreenYellow}
\mata             &\best 0 & \best{39} & \best \nearzero & 401 &\best 0 & \best{100} & \best{10} & 286 &\best 0 & \best \nearzero & \best \nearzero & 3 & \best{156} & \best{1\,s} & TO & 4\,s &\best 0 & \best{166} & \best 7 & 326\\
\awali            &\best 0 & 73 & 2 & 638 &\best 0 & 490 & 55 & 1\,s & 6 & 3 & 1 & 7 & 157 & 6\,s & TO & 7\,s &\best 0 & 1\,s & 81 & 3\,s\\
\vata             &\best 0 & 57 & 2 & 296 & \multicolumn{4}{c}{-} & 2 & 4 & \best \nearzero & 22 & 159 & 7\,s & TO & 8\,s & 14 & 6\,s & 270 & 12\,s\\
\automatanet      &\best 0 & 53 & 39 & 110 & \multicolumn{4}{c}{-} &\best 0 & 26 & 24 & 9 & 157 & 8\,s & TO & 10\,s &\best 0 & 220 & 47 & 314\\
\brics            &\best 0 & 47 & 8 & 190 &\best 0 & 136 & 35 & 204 &\best 0 & 7 & 3 & 21 & 159 & 6\,s & TO & 6\,s &\best 0 & 223 & 50 & 307\\
\automatajar      &\best 0 & 293 & 143 & 793 & \multicolumn{4}{c}{-} & 17 & 276 & 216 & 675 & 227 & 8\,s & TO & 13\,s & 227 & 10\,s & TO & 15\,s\\
\fado             & 10 & 646 & 5 & 4\,s & 189 & 10\,s & 25\,s & 13\,s & 10 & 271 & 52 & 2\,s & 250 & 15\,s & TO & 20\,s & 115 & 5\,s & 12\,s & 11\,s\\
\automatapy       & 3 & 263 & 5 & 2\,s & \multicolumn{4}{c}{-} & 5 & 38 & 3 & 353 & 254 & 4\,s & TO & 6\,s & 245 & 11\,s & TO & 16\,s\\
\bottomrule

\end{tabular}
}
\end{center}
\vspace*{-8mm}
\end{table}

\begin{table}[t]
\begin{center}
\caption{Relative speedup of \mata on instances where both libraries finished.}
\label{table:relative}
\newcolumntype{g}{>{\columncolor{gray!30}}r}
\newcolumntype{h}{>{\columncolor{gray!30}}c}

\begin{tabular}{l grgrgrg}
\toprule
               & \awali & \vata  & \automatanet   &  \brics & \automatajar  &   \multicolumn{1}{c}{\fado} & \automatapy   \\
\midrule
\bincl          &  27.52 & 1.86   & 29.73          &   16.98 & 21.44         & 4839.55 & 23.22         \\
\bcsmtlib       &   3.7  & 4.52   & 89.64          &   26.13 & 236.36        &   70.16 & 24.47         \\
\bregexlib      &  25.07 & 22.59  & 37.19          &   55.3  & 273.35        & 9999.29 & 282.41        \\
\bpresexpl      &   2.22 & 2.88   & 2.69           &    1.57 & 10.89         &   85.17 & 25.38         \\
\bpressym       &   3.46 & 4.65   & 51.82          &   27.99 & 82.47         &   67.54 & 17.97         \\
\bnoodlercompl  &   1.85 & 1.45   & 1.37           &    1.22 & 7.44          &  137.53 & 15.58         \\
\bnoodlerconc   &   4.87 & -      & -              &    1.36 & -             & 1979.56 & -             \\
\bnoodlerinter  &   4.02 & 6.42   & 33.98          &    9.04 & 371.23        &  363.49 & 51.51         \\
\bparam         &   5.36 & 7.3    & 7.27           &    6.49 & 1.43          & 2148.64 & 58.85         \\
\bparamunion    &   8.61 & 51.77  & 1.33           &    1.34 & 833.69        & 1618.04 & 5860.62       \\
\bottomrule
\end{tabular}

\end{center}
\end{table}

\begin{figure}[t]
\captionsetup{font=small}
\begin{center}
 \includegraphics[width=\linewidth]{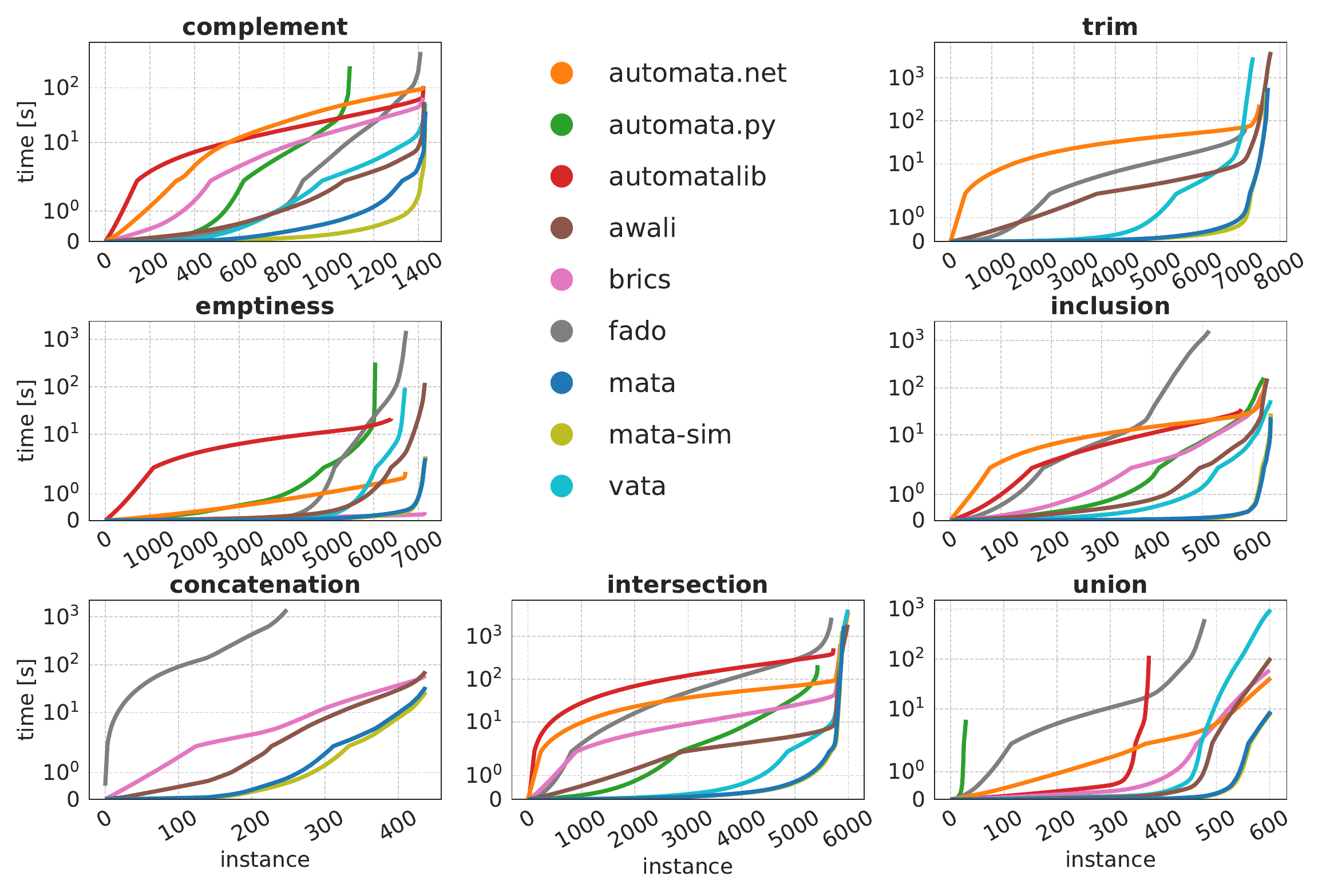}
\caption{Cactus plot showing cumulative run time per operation. The time axis is logarithmic. }
\label{fig:operations}
\end{center}
\vspace{-4mm}
\end{figure}

\paragraph{Benchmarks.}

We use the following benchmark sets.
\vspace{-2mm}
\begin{description}
\item[\bcsmtlib~\cite{cade23}]
contains 384 instances of boolean combinations of regular properties, obtained from SMT formulae over the theory of strings.
These include difficult handwritten problems containing membership in regular expressions extended with intersection and complement from~\cite{margus_derivatives_21} and emptiness problems from Norn~\cite{AutomataSplitting,norn} and SyGuS-qgen benchmarks, collected in SMT-LIB~\cite{BarFT-SMTLIB,QF_S,QF_SLIA}.

\item[\bregexlib~\cite{cade23}] contains 500 inclusion checks of the form $r_5 \subseteq r_1 \land r_2 \land r_3 \land r_4$ obtained analogously as in \cite{dantoni_afa_2016}. Each $r_i$ is one of the 75 regexes\footnote{\url{https://github.com/lorisdanto/symbolicautomata/blob/master/benchmarks/src/main/java/regexconverter/pattern\%4075.txt}} from RegExLib~\cite{regexlib}, selected so that $r_1 \land r_2 \land r_3 \land r_4 \land r_5$ is not empty.
Similar kind of these problems is solved in spam-filtering: one tests whether a new filter $r_5$ adds anything new to existing filters.

\item[\bparam~\cite{cade23}]
contains 4 sets of parametric intersection problems from~\cite{gange_unbounded_13} and 2 sets from \cite{cox_paper_17}. In total, this includes 267 problems.
The parameter controls the size of the regex or the number of regexes to be combined. 
\bparamunion is the variant of the benchmark that performs union instead of intersection.

\item[\bincl~\cite{cade23}]
contains 136 language inclusion problems derived from runs of an abstract regular model checker of \cite{bouajjani-antichain-08} (verification of the bakery algorithm, bubble sort, and a pro\-duc\-er-con\-sumer system).

\item[
 \benchname{lia}
]
consists of 169 complementation problems created during the run of Amaya~\cite{amaya}, a~tool for deciding linear integer arithmetic (LIA) formulae
using an automata-based decision procedure of \cite{BoudetComon}.
The formulae are taken from \smtbenchfont{UltimateAutomizer}~\cite{UltimateAutomizer} and \smtbenchfont{tptp}~\cite{tptp} benchmarks, collected in SMT-LIB~\cite{BarFT-SMTLIB,LIA}.
The transition relation in Amaya is represented symbolically using BDDs; in our experiments we tested both symbolic representation (in \bpressym) and explicit representation (in \bpresexpl), where explicit symbols are bit vectors represented by the BDDs.

\item[\benchname{noodler}]
consists of instances created during the run of the string solver \noodler \cite{spagheti23,ChenCHHLS23,noodlertool23} on the regex-heavy benchmark AutomatArk~\cite{Z3str3RE} from SMT-LIB~\cite{BarFT-SMTLIB,QF_S}. We collected 751 complementation, 438 concatenation, and 4,872 intersection problems in \bnoodlercompl, \bnoodlerconc, and \bnoodlerinter respectively.
\end{description}

\paragraph{Experimental setup.}
We converted all benchmarks into a common textual automata format 
(the \code{.mata} format, see \cref{sec:infrastructure}), 
and wrote dedicated parsers or conversions for all the libraries.
The conversion and parsing are not included in the run times since the parsers are not optimized and the typical use cases do not require parsing every input automaton from a textual format. From some of the benchmarks, we excluded small units of examples where the conversion failed.
\chck{We measure only the time needed for carrying out the specified operations on automata already parsed into  each library's internal data structures.}
%
Automata in all benchmarks but \bpres and those coming from regexes, \bregexlib, \bcsmtlib, \bparam, and \bparamunion, had small or moderate alphabet sizes (all below 100 symbols, except \bnoodlerinter with up to 252 symbols).
The explicit automata from LIA solving (\bpresexpl) have at most 1,024 symbols (corresponding to 10 bits).\footnote{It should be noted that these LIA problems are by no means representative of typical LIA formulae, which could generate much larger alphabets and transition relations that require some sort of symbolic representation.}
After performing mintermization on automata with symbolic representation (\bpressym), the number of symbols was reduced to at most 30,
and mintermization runs on automata from regular expressions 
returned alphabets with at most 80 symbols. 
\paragraph{Results.}
We summarize the results of each benchmark in cactus plots in \cref{fig:benchmarks} (displaying cumulative run times of benchmarks, with the instances ordered by their run time) and \cref{table:benchmarks}. \Cref{table:relative} shows relative speedups of \mata over each library on problem instances that both libraries finished in time.  
We also present statistics for individual automata operations across the entire benchmark in \cref{fig:operations} and \cref{table:operations}.
We do not show the performance of \mata's Python interface in the plots and tables as it is matches that one of \mata. 
All examples were run in six parallel jobs on Fedora GNU/Linux 38 with an Intel Core 3.4\,GHz processor and 20\,GiB RAM with 60\,s timeout. 

\mata consistently outperforms all other libraries on all benchmarks and in all operations, up to few exceptions. It is sometimes matched or outperformed by \automatanet and \brics in union and concatenation operation (on \bparamunion and \bnoodlerconc). 
\brics and \automatanet are sometimes faster since they may be able to share parts of the representation (such as BDDs on the transitions) between the automata operands and the union/concatenation, while \mata copies the entire data structure (and the memory locality of \deltastruct, with its three layers of vectors, is not perfect). 
\brics appears particularly fast in emptiness checking since it implicitly trims the automata, after which the emptiness test becomes a trivial query on emptiness of the set of states. 
The cost of the emptiness check is thus hidden in the cost of other operations (we do not state statistics from trimming for \brics for this reason). 
\brics and \automatanet also have a smaller average time in constructing the complements in $\bpressym$, due to a few high run times of \mata on examples that have many transitions per a pair of states. Solving these examples, and generally examples generated from solving LIA, is indeed a case for symbolic representation of transitions, and it is currently not a primary target of \mata. However, \mata is still much faster than any other library on mintermised versions of the same examples. 
\automatajar is faster in some parametric intersection examples because of its implicit determinization, which in some particular examples returns much smaller automata. When the other libraries are made to determinize, they behave analogously, and \mata again solves most examples and takes the least time. 
%
%
%
Still, on all operations except emptiness, \mata is the fastest overall, and on emptiness it is by far the fastest from libraries that actually do solve the emptiness problem. 
%
\mata has especially efficient inclusion test, and trimming, an operation which is usually needed very frequently, is also a strong point of \mata's performance. 

{\mata}'s simulation reduction (\matasim in the results) does not help much when the time for computing the simulation is counted in, as seen in \cref{fig:benchmarks}. 
Simulation reduction is indeed costly, and our eager strategy of reducing all automata is probably sub-optimal. The run times of complement, however, show a~considerable speedup after automata are reduced, and \matasim solves some complement and also parametric intersection examples that no other library can. 

Overall, \mata appears significantly faster than all the libraries we have tried, with the closest competitor being often more than an order of magnitude slower.

\begin{table}[t]
\caption{Statistics for the operations on solved instances. 
We list the average time (\textit{Avg}), median time (\textit{Med}), and standard deviation (\textit{Std}), with the best values in \textbf{bold}. 
The times are in milliseconds. 
Note that only the operations that the given library finished within the timeout are counted, hence the numbers are significantly biased in favour of libraries that timeouted more (the harder benchmarks are no counted in), and should be red in the context of \cref{table:benchmarks} and the cactus plots. 
We use \nearzero to denote a~value close to zero.
}
\label{table:operations}

\newcolumntype{g}{>{\columncolor{gray!30}}r}
\newcolumntype{h}{>{\columncolor{gray!30}}c}
\begin{center}
\resizebox{\textwidth}{!}{%
\begin{tabular}{l ggg rrr ggg rrr ggg rrr ggg}
 \toprule
 & \multicolumn{3}{h}{\textbf{complement}} & \multicolumn{3}{c}{\textbf{concatenation}} & \multicolumn{3}{h}{\textbf{emptiness}} & \multicolumn{3}{c}{\textbf{inclusion}} & \multicolumn{3}{h}{\textbf{intersection}} & \multicolumn{3}{c}{\textbf{trim}} & \multicolumn{3}{h}{\textbf{union}}\\\cmidrule(lr){2-4}\cmidrule(lr){5-7}\cmidrule(lr){8-10}\cmidrule(lr){11-13}\cmidrule(lr){14-16}\cmidrule(lr){17-19}\cmidrule(lr){20-22}
& \multicolumn{1}{h}{\textit{Avg}} & \multicolumn{1}{h}{\textit{Med}} & \multicolumn{1}{h}{\textit{Std}} & \multicolumn{1}{c}{\textit{Avg}} & \multicolumn{1}{c}{\textit{Med}} & \multicolumn{1}{c}{\textit{Std}} & \multicolumn{1}{h}{\textit{Avg}} & \multicolumn{1}{h}{\textit{Med}} & \multicolumn{1}{h}{\textit{Std}} & \multicolumn{1}{c}{\textit{Avg}} & \multicolumn{1}{c}{\textit{Med}} & \multicolumn{1}{c}{\textit{Std}} & \multicolumn{1}{h}{\textit{Avg}} & \multicolumn{1}{h}{\textit{Med}} & \multicolumn{1}{h}{\textit{Std}} & \multicolumn{1}{c}{\textit{Avg}} & \multicolumn{1}{c}{\textit{Med}} & \multicolumn{1}{c}{\textit{Std}} & \multicolumn{1}{h}{\textit{Avg}} & \multicolumn{1}{h}{\textit{Med}} & \multicolumn{1}{h}{\textit{Std}}\\
\midrule
\rowcolor{GreenYellow}
\mata             & \best{25} & \best 1 & 315 & \best{78} & \best 8 & 235 & \best \nearzero & \best \nearzero & 2 & \best{37} & \best \nearzero & 576 & 295 & \best \nearzero & 3\,s & 76 & \best \nearzero & 828 & \best{14} & \best \nearzero & 45\\
\awali            & 38 & 2 & 462 & 166 & 22 & 402 & 17 & \best \nearzero & 138 & 250 & 2 & 2\,s & 312 & \best \nearzero & 2\,s & 516 & \best \nearzero & 4\,s & 173 & \best \nearzero & 527\\
\vata             & 36 & 3 & 294 & \multicolumn{3}{c}{-} & 14 & \best \nearzero & 130 & 85 & 1 & 374 & 699 & \best \nearzero & 4\,s & 408 & \best \nearzero & 3\,s & 2\,s & \best \nearzero & 5\,s\\
\automatanet      & 73 & 59 & 89 & \multicolumn{3}{c}{-} & \best \nearzero & \best \nearzero & \nearzero & 245 & 43 & 1\,s & 621 & 14 & 4\,s & 31 & 9 & 165 & 69 & 6 & 163\\
\brics            & 46 & 24 & 140 & 136 & 35 & 204 & \best \nearzero & \best \nearzero & \nearzero & 204 & 10 & 1\,s & 115 & 4 & 1\,s & \multicolumn{3}{c}{-} & 99 & 2 & 232\\
\automatajar      & 75 & 31 & 657 & \multicolumn{3}{c}{-} & 3 & 2 & 5 & 60 & 42 & 102 & 91 & 59 & 748 & \multicolumn{3}{c}{-} & 311 & 2 & 3\,s\\
\fado             & 320 & 3 & 2\,s & 6\,s & 10\,s & 10\,s & 223 & \best \nearzero & 2\,s & 3\,s & 84 & 8\,s & 479 & 48 & 3\,s & \best{10} & 3 & 70 & 1\,s & 84 & 6\,s\\
\automatapy       & 226 & 25 & 2\,s & \multicolumn{3}{c}{-} & 53 & \best \nearzero & 1\,s & 263 & 6 & 1\,s & \best{39} & 2 & 479 & \multicolumn{3}{c}{-} & 203 & TO & 377\\
\bottomrule

\end{tabular}
}
\end{center}
\end{table}

\paragraph{Threats to validity.} Our results must be taken with a grain of salt
as the experiment contains an inherent room for error. 
Mainly, not knowing every library intimately,
we might have missed the most optimal solutions,
and our parsers of the \code{.mata} format 
might be building the internal data structures of the libraries in a sub-optimal way. 
The experiment was also running in parallel on a server with limited resources,
which might lead to fluctuations in run times
We are, however, confident that our main conclusions are well justified.

\section{Conclusions and Future Work}
We have introduced a new automata library \mata, explained its principles, and evaluated its performance. 
\mata is not the most general or feature-full library. Libraries such as \awali or \automatanet are much more complex and comprehensive, are more widely applicable, either to various symbolic representations of automata or to automata with registers, while still being impressively efficient.
\mata, however, does what it is meant to do better than all the other libraries: solve examples from string solving, regular expression processing, and regular model checking much faster, while staying simple and transparent, easily extensible and applicable to projects.


We continue working on \mata's set of features as well as its efficiency.
We plan to extend \mata with transducers, add support for registers that could handle, e.g., counting in regular expressions, and experiment with the poor man's symbolic representation of bit vector alphabets represented as sequences of bits (used in \lash~\cite{lash}), so that \mata can be used adequately in applications such as solving WS1S and arithmetic formulae.
We believe that the efficiency of the basic data structures discussed here can be much improved by focusing on the low-level performance. Custom data structures, specialised memory management, improvement in memory locality, and, generally, the class of optimizations used in BDD packages, could shift \mata's performance much further.

\section*{Acknowledgments}
This work has been supported by 
the Czech Ministry of Education, Youth and Sports ERC.CZ project LL1908,
the Czech Science Foundation project 23-07565S, and
the FIT BUT internal project FIT-S-23-8151.

\vspace{-0.0mm}
\section*{Data Availability Statement}\label{sec:label}
\vspace{-0.0mm}

An environment with the tools and data used for the experimental evaluation in
the current study is available at~\cite{mataArtifact}.

\bibliographystyle{splncs04}
\bibliography{literature.bib}

\begin{thebibliography}{10}
\providecommand{\url}[1]{\texttt{#1}}
\providecommand{\urlprefix}{URL }
\providecommand{\doi}[1]{https://doi.org/#1}

\bibitem{Trau}
Abdulla, P.A., Atig, M.F., Chen, Y., Diep, B.P., Hol{\'{\i}}k, L., Rezine, A.,
  R{\"{u}}mmer, P.: Trau: {SMT} solver for string constraints. In: Proc. of
  {FMCAD}'18. {IEEE} (2018)

\bibitem{AutomataSplitting}
Abdulla, P.A., Atig, M.F., Chen, Y., Hol{\'{\i}}k, L., Rezine, A.,
  R{\"{u}}mmer, P., Stenman, J.: String constraints for verification. In:
  Computer Aided Verification - 26th International Conference, {CAV} 2014, Held
  as Part of the Vienna Summer of Logic, {VSL} 2014, Vienna, Austria, July
  18-22, 2014. Proceedings. Lecture Notes in Computer Science, vol.~8559, pp.
  150--166. Springer (2014). \doi{10.1007/978-3-319-08867-9\_10},
  \url{https://doi.org/10.1007/978-3-319-08867-9\_10}

\bibitem{norn}
Abdulla, P.A., Atig, M.F., Chen, Y.F., Hol{\'i}k, L., Rezine, A., R{\"u}mmer,
  P., Stenman, J.: Norn: An {SMT} solver for string constraints. In: Computer
  Aided Verification. pp. 462--469. Springer International Publishing, Cham
  (2015)

\bibitem{treesimulation08}
Abdulla, P.A., Bouajjani, A., Hol{\'{\i}}k, L., Kaati, L., Vojnar, T.:
  Computing simulations over tree automata. In: Ramakrishnan, C.R., Rehof, J.
  (eds.) Tools and Algorithms for the Construction and Analysis of Systems,
  14th International Conference, {TACAS} 2008, Held as Part of the Joint
  European Conferences on Theory and Practice of Software, {ETAPS} 2008,
  Budapest, Hungary, March 29-April 6, 2008. Proceedings. Lecture Notes in
  Computer Science, vol.~4963, pp. 93--108. Springer (2008).
  \doi{10.1007/978-3-540-78800-3\_8},
  \url{https://doi.org/10.1007/978-3-540-78800-3\_8}

\bibitem{abdulla_when_2010}
Abdulla, P.A., Chen, Y.F., Hol{\'i}k, L., Mayr, R., Vojnar, T.: When simulation
  meets antichains. In: Proc. of TACAS'10. LNCS, vol.~6015. Springer (2010)

\bibitem{rmc_survey04}
Abdulla, P.A., Jonsson, B., Nilsson, M., Saksena, M.: A survey of regular model
  checking. In: Gardner, P., Yoshida, N. (eds.) CONCUR 2004 - Concurrency
  Theory. pp. 35--48. Springer Berlin Heidelberg, Berlin, Heidelberg (2004)

\bibitem{fado}
Almeida, A., Almeida, M., Alves, J., Moreira, N., Reis, R.: Fado and guitar:
  Tools for automata manipulation and visualization. In: Maneth, S. (ed.)
  Implementation and Application of Automata. pp. 65--74. Springer Berlin
  Heidelberg, Berlin, Heidelberg (2009)

\bibitem{amaya}
authors, A.: Amaya (2023), \url{https://github.com/MichalHe/amaya}

\bibitem{BarFT-SMTLIB}
Barrett, C., Fontaine, P., Tinelli, C.: {The Satisfiability Modulo Theories
  Library ({SMT-LIB})}. {\tt www.SMT-LIB.org} (2016)

\bibitem{Z3str3RE}
Berzish, M., Kulczynski, M., Mora, F., Manea, F., Day, J.D., Nowotka, D.,
  Ganesh, V.: An {SMT} solver for regular expressions and linear arithmetic
  over string length. In: Computer Aided Verification - 33rd International
  Conference, {CAV} 2021, Virtual Event, July 20-23, 2021, Proceedings, Part
  {II}. Lecture Notes in Computer Science, vol. 12760, pp. 289--312. Springer
  (2021). \doi{10.1007/978-3-030-81688-9\_14},
  \url{https://doi.org/10.1007/978-3-030-81688-9\_14}

\bibitem{spagheti23}
Blahoudek, F., Chen, Y.F., Chocholat{\'y}, D., Havlena, V., Hol{\'i}k, L.,
  Leng{\'a}l, O., S{\'i}{\v{c}}, J.: Word equations in synergy with regular
  constraints. In: Proc. of FM'23. Springer (2023)

\bibitem{lash}
Boigelot, B., Latour, L.: Counting the solutions of {Presburger} equations
  without enumerating them. Theoretical Computer Science  \textbf{313}(1),
  17--29 (2004). \doi{https://doi.org/10.1016/j.tcs.2003.10.002},
  \url{https://www.sciencedirect.com/science/article/pii/S0304397503005322},
  implementation and Application of Automata

\bibitem{iterating_transd03}
Boigelot, B., Legay, A., Wolper, P.: Iterating transducers in the large. In:
  Hunt, W.A., Somenzi, F. (eds.) Computer Aided Verification. pp. 223--235.
  Springer Berlin Heidelberg, Berlin, Heidelberg (2003)

\bibitem{bonchi_checking_2013}
Bonchi, F., Pous, D.: Checking {NFA} equivalence with bisimulations up to
  congruence. In: Proc. of {POPL}'13. {{ACM}} (2013)

\bibitem{bouajjani-antichain-08}
Bouajjani, A., Habermehl, P., Hol{\'i}k, L., Touili, T., Vojnar, T.:
  Antichain-based universality and inclusion testing over nondeterministic
  finite tree automata. In: Proc. of CIAA'08. Springer (2008)

\bibitem{armc04}
Bouajjani, A., Habermehl, P., Vojnar, T.: Abstract regular model checking. In:
  Alur, R., Peled, D.A. (eds.) Computer Aided Verification, 16th International
  Conference, {CAV} 2004, Boston, MA, USA, July 13-17, 2004, Proceedings.
  Lecture Notes in Computer Science, vol.~3114, pp. 372--386. Springer (2004).
  \doi{10.1007/978-3-540-27813-9\_29},
  \url{https://doi.org/10.1007/978-3-540-27813-9\_29}

\bibitem{BoudetComon}
Boudet, A., Comon, H.: Diophantine equations, {Presburger} arithmetic and
  finite automata. In: Kirchner, H. (ed.) Trees in Algebra and Programming ---
  CAAP '96. pp. 30--43. Springer Berlin Heidelberg, Berlin, Heidelberg (1996)

\bibitem{sparseset93}
Briggs, P., Torczon, L.: An efficient representation for sparse sets. ACM Lett.
  Program. Lang. Syst.  \textbf{2}(1–4),  59–69 (mar 1993).
  \doi{10.1145/176454.176484}, \url{https://doi.org/10.1145/176454.176484}

\bibitem{Brzozowski1962CanonicalRE}
Brzozowski, J.A.: Canonical regular expressions and minimal state graphs for
  definite events. In: Proc. of Symposium on Mathematical Theory of Automata
  (1962)

\bibitem{Buchi1990}
B{\"u}chi, J.R.: Weak Second-Order Arithmetic and Finite Automata, pp.
  398--424. Springer New York, New York, NY (1990).
  \doi{10.1007/978-1-4613-8928-6_22},
  \url{https://doi.org/10.1007/978-1-4613-8928-6_22}

\bibitem{Cece17}
C{\'{e}}c{\'{e}}, G.: Foundation for a series of efficient simulation
  algorithms. In: Proc. of {LICS'17}. {IEEE} (2017)

\bibitem{AnthonyReplaceAll2018}
Chen, T., Chen, Y., Hague, M., Lin, A.W., Wu, Z.: What is decidable about
  string constraints with the replaceall function. Proc. of {POPL}'18  (2018)

\bibitem{AnthonyComplex2019}
Chen, T., Hague, M., Lin, A.W., R{\"{u}}mmer, P., Wu, Z.: Decision procedures
  for path feasibility of string-manipulating programs with complex operations.
  Proc. of POPL'19  (2019)

\bibitem{ChenCHHLS23}
Chen, Y.F., Chocholat\'{y}, D., Havlena, V., Hol\'{\i}k, L., Leng\'{a}l, O.,
  S\'{\i}\v{c}, J.: Solving string constraints with lengths by stabilization.
  Proc. ACM Program. Lang.  \textbf{7}(OOPSLA2) (oct 2023).
  \doi{10.1145/3622872}

\bibitem{noodlertool23}
Chen, Y.F., Chocholatý, D., Havlena, V., Holík, L., Lengál, O., Síč, J.:
  Z3-noodler: An automata-based string solver. In: Proc. of TACAS'24. LNCS,
  Springer (2024)

\bibitem{learningrmc17}
Chen, Y., Hong, C., Lin, A.W., R{\"{u}}mmer, P.: Learning to prove safety over
  parameterised concurrent systems. In: Stewart, D., Weissenbacher, G. (eds.)
  2017 Formal Methods in Computer Aided Design, {FMCAD} 2017, Vienna, Austria,
  October 2-6, 2017. pp. 76--83. {IEEE} (2017).
  \doi{10.23919/FMCAD.2017.8102244},
  \url{https://doi.org/10.23919/FMCAD.2017.8102244}

\bibitem{mataArtifact}
Chocholatý, D., Fiedor, T., Havlena, V., Holík, L., Hruška, M., Lengál, O.,
  Síč, J.: A replication package for reproducing the results of paper
  ``\mata: A fast and simple finite automata library'' (Oct 2023).
  \doi{10.5281/zenodo.10044515}, \url{https://doi.org/10.5281/zenodo.10044515}

\bibitem{cox_paper_17}
Cox, A., Leasure, J.: Model checking regular language constraints. CoRR
  \textbf{abs/1708.09073} (2017)

\bibitem{lorisjava}
D'Antoni, L.: {A symbolic automata library},
  \url{https://github.com/lorisdanto/symbolicautomata}

\bibitem{dantoni_afa_2016}
D'Antoni, L., Kincaid, Z., Wang, F.: A symbolic decision procedure for symbolic
  alternating finite automata. Electronic Notes in Theoretical Computer Science
   \textbf{336} (2018)

\bibitem{margus_minimization}
D'Antoni, L., Veanes, M.: Minimization of symbolic automata. In: Proc. of
  POPL'14. ACM (2014)

\bibitem{dantoni_taminimization_2016}
D'Antoni, L., Veanes, M.: Minimization of symbolic tree automata. In: Proc. of
  {LICS}'16. {{ACM}} (2016)

\bibitem{margus_the_power17}
D'Antoni, L., Veanes, M.: The power of symbolic automata and transducers. In:
  Majumdar, R., Kun{\v{c}}ak, V. (eds.) Computer Aided Verification. pp.
  47--67. Springer International Publishing, Cham (2017)

\bibitem{alaska}
De~Wulf, M., Doyen, L., Maquet, N., Raskin, J.F.: Alaska. In: Proc. of ATVA'08.
  Springer (2008)

\bibitem{doyen-antichain-10}
Doyen, L., Raskin, J.: Antichain algorithms for finite automata. In: Proc. of
  {TACAS}'10. LNCS, Springer (2010)

\bibitem{spot}
Duret-Lutz, A., Renault, E., Colange, M., Renkin, F., Gbaguidi~Aisse, A.,
  Schlehuber-Caissier, P., Medioni, T., Martin, A., Dubois, J., Gillard, C.,
  Lauko, H.: From {Spot} 2.0 to {Spot} 2.10: What's new? In: Shoham, S., Vizel,
  Y. (eds.) Computer Aided Verification. pp. 174--187. Springer International
  Publishing, Cham (2022)

\bibitem{automatapy}
Evans, C.: Automata (2023), \url{https://github.com/caleb531/automata}

\bibitem{cade23}
Fiedor, T., Hol{\'{\i}}k, L., Hruska, M., Rogalewicz, A., S{\'{\i}}c, J.,
  Vargov\v{c}{\'{\i}}k, P.: Reasoning about regular properties: {A} comparative
  study. In: Pientka, B., Tinelli, C. (eds.) Automated Deduction - {CADE} 29 -
  29th International Conference on Automated Deduction, Rome, Italy, July 1-4,
  2023, Proceedings. Lecture Notes in Computer Science, vol. 14132, pp.
  286--306. Springer (2023). \doi{10.1007/978-3-031-38499-8\_17},
  \url{https://doi.org/10.1007/978-3-031-38499-8\_17}

\bibitem{chenfu_eqchecking_17}
Fu, C., Deng, Y., Jansen, D.N., Zhang, L.: On equivalence checking of
  nondeterministic finite automata. In: Proc. of {SETTA}'17. LNCS, Springer
  (2017)

\bibitem{gange_unbounded_13}
Gange, G., Navas, J.A., Stuckey, P.J., S{\o}ndergaard, H., Schachte, P.:
  Unbounded model-checking with interpolation for regular language constraints.
  In: Proc. of {TACAS}'13. LNCS, Springer (2013)

\bibitem{re2}
Google: Re2. \url {https://github.com/google/re2}

\bibitem{UltimateAutomizer}
Heizmann, M., Hoenicke, J., Podelski, A.: Software model checking for people
  who love automata. In: Sharygina, N., Veith, H. (eds.) Computer Aided
  Verification. pp. 36--52. Springer Berlin Heidelberg, Berlin, Heidelberg
  (2013)

\bibitem{mona}
Henriksen, J.G., Jensen, J.L., J{\o}rgensen, M.E., Klarlund, N., Paige, R.,
  Rauhe, T., Sandholm, A.: Mona: {M}onadic second-order logic in practice. In:
  Proc. of {TACAS} '95. LNCS, vol.~1019. Springer (1995)

\bibitem{HHK95}
Henzinger, M.R., Henzinger, T.A., Kopke, P.W.: Computing simulations on finite
  and infinite graphs. In: Proc. of FOCS. {IEEE} (1995)

\bibitem{janku_string_2018}
Hol{\'{\i}}k, L., Jank{\r u}, P., Lin, A.W., R{\"{u}}mmer, P., Vojnar, T.:
  String constraints with concatenation and transducers solved efficiently.
  Proc. of {POPL}'18  \textbf{2} (2018)

\bibitem{symbsim18}
Hol{\'i}k, L., Leng{\'a}l, O., S{\'i}{\v{c}}, J., Veanes, M., Vojnar, T.:
  Simulation algorithms for symbolic automata. In: Lahiri, S.K., Wang, C.
  (eds.) Proc. of ATVA'18. Springer (2018)

\bibitem{tree_inclusion_11}
Hol{\'{\i}}k, L., Leng{\'{a}}l, O., {\v{S}}im{\'{a}}{\v{c}}ek, J., Vojnar, T.:
  Efficient inclusion checking on explicit and semi-symbolic tree automata. In:
  Proc. of {ATVA}'11. LNCS, Springer (2011)

\bibitem{lukasjirisimulation}
Hol\'{i}k, L., \v{S}im\'{a}\v{c}ek, J.: Optimizing an {LTS}-simulation
  algorithm. Computing and Informatics  \textbf{29}(6+),  1337--1348 (2010),
  \url{https://arxiv.org/abs/2307.04235}

\bibitem{dprle}
Hooimeijer, P., Weimer, W.: A decision procedure for subset constraints over
  regular languages. In: PLDI'09. ACM (2009)

\bibitem{hopcroft_71}
Hopcroft, J.E.: An n log n algorithm for minimizing states in a finite
  automaton. Tech. rep., Stanford University, Stanford, CA, USA (1971)

\bibitem{HUFFMAN1954}
Huffman, D.: The synthesis of sequential switching circuits. Journal of the
  Franklin Institute  \textbf{257}(3) (1954)

\bibitem{Ilie2004}
Ilie, L., Navarro, G., Yu, S.: On {NFA} reductions. In: Theory Is Forever:
  Essays Dedicated to Arto Salomaa on the Occasion of His 70th Birthday.
  Springer (2004)

\bibitem{automatajar}
Isberner, M., Howar, F., Steffen, B.: {AutomataLib},
  \url{https://learnlib.de/projects/automatalib/}

\bibitem{learnlib}
Isberner, M., Howar, F., Steffen, B.: The open-source learnlib. In: Kroening,
  D., P{\u{a}}s{\u{a}}reanu, C.S. (eds.) Computer Aided Verification. pp.
  487--495. Springer International Publishing, Cham (2015)

\bibitem{mosel}
Kelb, P., Margaria, T., Mendler, M., Gsottberger, C.: {MOSEL:} {A} sound and
  efficient tool for {M2L(Str)}. In: Grumberg, O. (ed.) Computer Aided
  Verification, 9th International Conference, {CAV} '97, Haifa, Israel, June
  22-25, 1997, Proceedings. Lecture Notes in Computer Science, vol.~1254, pp.
  448--451. Springer (1997). \doi{10.1007/3-540-63166-6\_45},
  \url{https://doi.org/10.1007/3-540-63166-6\_45}

\bibitem{Klaedtke2004}
Klaedtke, F.C.: Automata-based decision procedures for weak arithmetics. Ph.D.
  thesis, University of Freiburg, Freiburg im Breisgau, Germany (2004),
  \url{http://freidok.ub.uni-freiburg.de/volltexte/1439/index.html}

\bibitem{owl}
K{\v{r}}et{\'i}nsk{\'y}, J., Meggendorfer, T., Sickert, S.: Owl: A library for
  {$\omega$}-words, automata, and {LTL}. In: Lahiri, S.K., Wang, C. (eds.)
  Automated Technology for Verification and Analysis. pp. 543--550. Springer
  International Publishing, Cham (2018)

\bibitem{tormc12}
Legay, A.: {T(O)RMC}: A tool for ($\omega$)-regular model checking. In: Gupta,
  A., Malik, S. (eds.) Computer Aided Verification. pp. 548--551. Springer
  Berlin Heidelberg, Berlin, Heidelberg (2008)

\bibitem{vata}
Leng{\'{a}}l, O., {\v{S}}im{\'{a}}{\v{c}}ek, J., Vojnar, T.: {VATA:} {A}
  library for efficient manipulation of non-deterministic tree automata. In:
  Proc. of {TACAS}'12. LNCS, vol.~7214. Springer (2012)

\bibitem{awali}
Lombardy, S., Marsault, V., Sakarovitch, J.: Awali, a library for weighted
  automata and transducers (version 2.0) (2021), software available at
  {http://vaucanson-project.org/Awali/2.0/}

\bibitem{libfa}
Lutterkort, D.: libfa, \url{https://augeas.net/libfa/}

\bibitem{Moore1956}
Moore, E.F.: Gedanken-experiments on sequential machines. In: Automata Studies.
  Volume 34. Princeton University Press, Princeton (1956)

\bibitem{brics}
Møller, A., et~al.: Brics automata library,
  \url{https://www.brics.dk/automaton/}

\bibitem{piagetarjan_87}
Paige, R., Tarjan, R.E.: Three partition refinement algorithms. SIAM Journal on
  Computing  \textbf{16}(6) (1987)

\bibitem{ranzato_efficient_2010}
Ranzato, F., Tapparo, F.: An efficient simulation algorithm based on abstract
  interpretation. Information and Computation  \textbf{208},  1--22 (2010)

\bibitem{regexlib}
RegExLib.com: {The Internet's first Regular Expression Library}.
  {\url{http://regexlib.com/}}

\bibitem{QF_S}
{SMT-LIB}: \url{https://clc-gitlab.cs.uiowa.edu:2443/SMT-LIB-benchmarks/QF_S}
  (2023)

\bibitem{QF_SLIA}
{SMT-LIB}:
  \url{https://clc-gitlab.cs.uiowa.edu:2443/SMT-LIB-benchmarks/QF_SLIA} (2023)

\bibitem{LIA}
{SMT-LIB}: \url{https://clc-gitlab.cs.uiowa.edu:2443/SMT-LIB-benchmarks/LIA}
  (2023)

\bibitem{CUDD}
Somenzi, F.: {CUDD}: {CU} decision diagram package release 3.0.0 (2015)

\bibitem{margus_derivatives_21}
Stanford, C., Veanes, M., Bj{\o}rner, N.S.: Symbolic boolean derivatives for
  efficiently solving extended regular expression constraints. In: Proc. of
  {PLDI}'21. {ACM} (2021)

\bibitem{tptp}
Sutcliffe, G.: {The TPTP Problem Library and Associated Infrastructure. From
  CNF to TH0, TPTP v6.4.0}. Journal of Automated Reasoning  \textbf{59}(4),
  483--502 (2017)

\bibitem{Tarjan71}
Tarjan, R.E.: Depth-first search and linear graph algorithms (working paper).
  In: 12th Annual Symposium on Switching and Automata Theory, East Lansing,
  Michigan, USA, October 13-15, 1971. pp. 114--121. {IEEE} Computer Society
  (1971). \doi{10.1109/SWAT.1971.10},
  \url{https://doi.org/10.1109/SWAT.1971.10}

\bibitem{TozawaH03}
Tozawa, A., Hagiya, M.: {XML} schema containment checking based on
  semi-implicit techniques. In: Ibarra, O.H., Dang, Z. (eds.) Implementation
  and Application of Automata, 8th International Conference, {CIAA} 2003, Santa
  Barbara, California, USA, July 16-18, 2003, Proceedings. Lecture Notes in
  Computer Science, vol.~2759, pp. 213--225. Springer (2003).
  \doi{10.1007/3-540-45089-0\_20},
  \url{https://doi.org/10.1007/3-540-45089-0\_20}

\bibitem{goal}
Tsay, Y.K., Chen, Y.F., Tsai, M.H., Wu, K.N., Chan, W.C.: Goal: A graphical
  tool for manipulating b{\"u}chi automata and temporal formulae. In: Grumberg,
  O., Huth, M. (eds.) Tools and Algorithms for the Construction and Analysis of
  Systems. pp. 466--471. Springer Berlin Heidelberg, Berlin, Heidelberg (2007)

\bibitem{Valmari_10}
Valmari, A.: Simple bisimilarity minimization in {O}(m log n) time. Fundamenta
  Informaticae  \textbf{105}(3) (2010)

\bibitem{automatanet}
Veanes, M.: A .{NET} automata library,
  \url{https://github.com/AutomataDotNet/Automata}

\bibitem{margus_rex_10}
Veanes, M., de~Halleux, P., Tillmann, N.: Rex: {S}ymbolic regular expression
  explorer. In: Proc. of {ICST}'10. {IEEE} (2010)

\bibitem{fangyu_circuit_16}
Wang, H., Tsai, T., Lin, C., Yu, F., Jiang, J.R.: String analysis via automata
  manipulation with logic circuit representation. In: Proc. of {CAV'16}. LNCS,
  vol.~9779. Springer (2016)

\bibitem{WolperB95}
Wolper, P., Boigelot, B.: An automata-theoretic approach to {Presburger}
  arithmetic constraints (extended abstract). In: Mycroft, A. (ed.) Proc. of
  SAS'95. LNCS, vol.~983. Springer (1995)

\bibitem{wolperrmc98}
Wolper, P., Boigelot, B.: Verifying systems with infinite but regular state
  spaces. In: Hu, A.J., Vardi, M.Y. (eds.) Computer Aided Verification. pp.
  88--97. Springer Berlin Heidelberg, Berlin, Heidelberg (1998)

\bibitem{dewulf_antichains_2006}
Wulf, M.D., Doyen, L., Henzinger, T.A., Raskin, J.: Antichains: {A} new
  algorithm for checking universality of finite automata. In: Proc. of
  {CAV}'06. LNCS, vol.~4144. Springer (2006)

\end{thebibliography}



\end{document}